{Review}

# Technology dictates algorithms: Recent developments in read alignment

Mohammed Alser[1,2,†], Jeremy Rotman[3,†], Kodi Taraszka[3], Huwenbo Shi[4,5], Pelin Icer Baykal[6],

Harry Taegyun Yang[3,7], Victor Xue[3], Sergey Knyazev[6], Benjamin D. Singer[8,9,10], Brunilda

Balliu[11], David Koslicki[12,13,14], Pavel Skums[6], Alex Zelikovsky[6,15], Can Alkan[2,16], Onur Mutlu[1,#],

Serghei Mangul[17,#,*]

[1]Computer Science Department, ETH Zürich, 8092 Zürich, Switzerland

[2]Computer Engineering Department, Bilkent University, 06800 Bilkent, Ankara, Turkey

[3]Department of Computer Science, University of California Los Angeles, Los Angeles, CA
90095, USA

[4]Department of Epidemiology, Harvard T.H. Chan School of Public Health, MA 02115, USA

[5]Program in Medical and Population Genetics, Broad Institute of MIT and Harvard, Cambridge,
MA 02142, USA

[6]Department of Computer Science, Georgia State University, Atlanta, GA 30302, USA

[7]Bioinformatics Interdepartmental Ph.D. Program, University of California Los Angeles, Los
Angeles, CA 90095, USA

[8]Division of Pulmonary and Critical Care Medicine, Northwestern University Feinberg School of
Medicine, Chicago, IL 60611, USA

[9]Department of Biochemistry & Molecular Genetics, Northwestern University Feinberg School
of Medicine, USA

[10]Simpson Querrey Center for Epigenetics, Northwestern University Feinberg School of Medicine, Chicago, IL 60611, USA

[11]Department of Computational Medicine, University of California Los Angeles, Los Angeles, CA 90095, USA

[12]Computer Science and Engineering, Pennsylvania State University, University Park, PA 16801, USA

[13]Biology Department, Pennsylvania State University, University Park, PA 16801, USA

[14]The Huck Institutes of the Life Sciences, Pennsylvania State University, University Park, PA 16801, USA

[15]The Laboratory of Bioinformatics, I.M. Sechenov First Moscow State Medical University, Moscow, 119991, Russia

[16]Bilkent-Hacettepe Health Sciences and Technologies Program, Ankara, Turkey

[17]Department of Clinical Pharmacy, School of Pharmacy, University of Southern California, Los Angeles, CA 90089, USA

† These authors contributed equally to this work

# These authors jointly supervised this work.

* Correspondence: serghei.mangul@gmail.com

**Abstract**

Massively parallel sequencing techniques have revolutionized biological and medical sciences by providing unprecedented insight into the genomes of humans, animals, and microbes. Modern sequencing platforms generate enormous amounts of genomic data in the form of nucleotide sequences or *reads*. Aligning reads onto reference genomes enables the identification of individual-specific genetic variants and is an essential step of the majority of genomic analysis pipelines. Aligned reads are essential for answering important biological questions, such as detecting mutations driving various human diseases and complex traits as well as identifying species present in metagenomic samples. The read alignment problem is extremely challenging due to the large size of analyzed datasets and numerous technological limitations of sequencing platforms, and researchers have developed novel bioinformatics algorithms to tackle these difficulties. Importantly, computational algorithms have evolved and diversified in accordance with technological advances, leading to today's diverse array of bioinformatics tools. Our review provides a survey of algorithmic foundations and methodologies across 107 alignment methods published between 1988 and 2020, for both short and long reads. We provide rigorous experimental evaluation of 11 read aligners to demonstrate the effect of these underlying algorithms on speed and efficiency of read aligners. We separately discuss how longer read lengths produce unique advantages and limitations to read alignment techniques. We also discuss how general alignment algorithms have been tailored to the specific needs of various domains in biology, including whole transcriptome, adaptive immune repertoire, and human microbiome studies.

## Introduction

In April 2003, the high-throughput sequencing era started with the Human Genome Project, which led to the successful sequencing of a nearly complete human genome and establishment of a reference genome that is still in use[1]. The Human Genome Project cost approximately $3 billion over 13 years to sequence the genome of an individual human. Recent advances in high-throughput sequencing technologies have enabled cost-effective and time-efficient probing of the DNA sequences of living organisms through a process known as DNA sequencing[2]. Modern high-throughput sequencing techniques are capable of producing millions of nucleotide sequences of an individual's DNA[3] and providing multifold coverage of whole genomes or particular genomic regions. The output of high-throughput sequencing consists of sets of relatively short genomic sequences, usually referred to as *reads*. Contemporary sequencing technologies are capable of generating tens of millions to billions of reads per sample, with read lengths ranging from a few hundred to a few million base pairs[4].

The trade-off for decreased cost and increased throughput offered by modern sequencing technologies is a larger margin of noise in sequencing data[5]. The magnitude of error rates in data produced by state-of-the-art sequencing platforms varies from $\sim10^{-3}$ for short reads to $\sim15\times10^{-2}$ for the relatively new long and ultra-long reads[6]. The increased error rate of today's emerging long read technologies may negatively impact biological interpretations. For example, errors in protein-coding regions can bias the accuracy of protein predictions[7]. Sequenced reads lack information about the order and origin (i.e., which part, homolog, and strand of the subject genome) of reads. The main challenge in genome analysis today is to reconstruct the complete genome of an individual. This process, *read alignment* (also known as *read mapping*), typically requires the reference genome which is used to determine the potential location of each read. Accuracy of alignment has a strong effect on many downstream analyses[8]. For example, most trans-eQTL signals were shown to be solely caused by alignment errors[9].

Read alignment can be performed in a brute force manner, which involves scanning the entire genome for the best matching portions to the read. The brute force approach is computationally expensive as it requires checking more than 3 billion positions in the human genome for alignment. Modern sequencing platforms are capable of producing hundreds of millions of reads, making brute force search infeasible in practice. Instead, today's efficient bioinformatics algorithms enable fast and accurate read alignment and can be thousands of orders of magnitude faster when compared to the naive brute force approach[10] (Supplementary Note 1).

Read alignment enables observation of the differences between the read and the reference genome. These differences can be caused by either real genetic variants in the sequenced genome or errors generated by the sequencing platform. These sequencing errors and read lengths, which are typically short, make the read alignment problem computationally challenging. The continued increase in the throughput of modern sequencing technologies creates additional demand for efficient algorithms for read alignment. Over the past several decades, a plethora of tools were developed to align reads onto reference genomes across various domains of biology. Previous efforts that provide overviews of various algorithms and techniques used by read aligners are presented elsewhere [10–12], including studies that present benchmarks of existing tools[13,14]. Since the time those efforts were published, many new alignment algorithms have been developed. Additionally, previous efforts lack a historical perspective on algorithm development.

Our review provides a historical perspective on how technological advancements in sequencing are shaping algorithm development across various domains of modern biology, and we systematically assess the underlying algorithms of a large number of aligners ($n = 107$). Algorithmic development and challenges associated with read alignment are to a large degree data- and technology-driven, and emerging highly-accurate ultra-long read sequencing techniques promise to expand the application of read alignment.

**Where do reads come from—advantages and limitations of read alignment**

One can study an individual genome using sequencing data in two ways: by mapping reads to a reference genome, if it exists, or by *de novo* assembling the reads. The complexity of the human genome, in combination with the short length of sequenced reads, poses substantial challenges to our ability to accurately assemble personal genomes[15]. Even recently-introduced ultra-long reads[16] (up to 2Mb) offer limited capacity to build a *de novo* assembly of an individual genome with no prior knowledge about the reference genome[16]. The presence of many repetitive regions in the human genome limits our ability to assemble a personal human genome as a single sequence. Emerging long read sequencing technologies that are capable of producing ultra-long reads[16] promise to deliver more accurate assemblies[17]. However, the relatively high error rate of data output from recently developed long read sequencing technologies often results in inaccuracies in the assembled genomes, especially when using low sequencing coverage[18,19].

The read alignment problem is known to be solvable in polynomial time[20], while a polynomial time solution for genome assembly is still unknown[20–22]. Genome assembly is typically slower and more computationally intensive than read alignment[17,23,24] due to the presence of repeats that are much longer than the typical read length. This makes assembly impractical in studies that involve large-scale clinical cohorts of thousands of individuals. At the same time, when the reference genome is unknown, long reads are a valuable resource for assembling genomes that are far more complex than the human genome, such as the hexaploid bread wheat genome[17,23,25]. Read alignment and genome assembly can be combined for so-called 'reference-guided *de novo* assembly', where reads compatible with the current reference are identified and remaining reads are assembled[26]. Similar approaches have been used to guide the assembly of genomes from related species[27].

The availability of a large number of software tools that are scalable to both read length and genome size have enabled read alignment to become an essential component of high throughput sequencing analysis (Table 1)[28]. However, read alignment also has its own fundamental challenges. First, some challenges are caused by the incompleteness of the reference genomes that have multiple assembly gaps[16]. Reads originating from these gaps often remain unmapped or are incorrectly mapped to homologous regions. Second, the presence of repetitive regions of the genome confounds current read alignment techniques, which often map reads originating from one region to match *several* other repetitive regions (such reads are known as multi-mapped reads). In such cases, most read aligners *simply* report one location randomly selected among the possible mapping locations, in turn, significantly reducing the number of detected variants[29].

Third, read alignment techniques should tolerate differences between reads and the reference genome. These differences may correspond to a single nucleotide (including deletion, insertion, and substitution of a nucleotide) or to larger structural variants[30]. Fourth, read alignment algorithms need to align reads to both forward and reverse DNA strands of the same reference genome in order to tackle the strand bias problem, defined as the difference in genotypes identified by reads that map to forward and reverse DNA strands. Strand bias is likely caused by errors introduced during library preparation and not by mapping artifacts[29,31].

Different areas of biological research pose additional, unique challenges to accurate genome alignment. For example, in viral biology studies, samples are often derived from unknown strains and may produce large differences when reads are aligned to known reference strains[32]. This necessitates the need for developing and using a new set of methods that, for example, represent genomic sequences as a graph instead of the linear representation of these sequences. The shared subsequences between the sequences are stored only once in the graph, and the differences between the sequences form branches in the graph. The graph-based approach supports bi-directional traversals, providing a more compact and comprehensive representation of genetic variations across many samples. Such methods require the development of novel sequence-to-graph alignment algorithms [33,34].

**Co-evolution of read alignment algorithms and sequencing technologies**

Over the past few decades, we have observed an increase in the number of alignment tools developed to accommodate rapid changes in sequencing technology (Table 1). Published alignment tools use a variety of algorithms to improve the accuracy and speed of read alignment (Table 2). At the same time, the development of read alignment algorithms are impacted by rapid changes in sequencing technologies, such as read length, throughput, and error rates (Supplementary Table 1). For example, some of the first alignment algorithms (e.g., BLAT[35]) were designed to align expressed sequence tag (EST) sequences, which are 200 to 500 bp in length. Another early alignment algorithm, BLASTZ[36], was designed to align 1Mb human contigs onto the mouse genome. After short reads became available, the majority of the algorithms have focused on the problem of aligning hundreds of millions of short reads to a reference genome. Recent sequencing technologies are capable of producing multi-megabase reads at the cost of high error rates (up to 20%)—a development that poses additional challenges for modern read alignment methods[17]. A recent improvement in circular consensus sequencing (CCS) allows substantial reduction in sequencing error rates; for example, the error rate has dropped from 15% down to 0.0001% by sequencing the same molecule at least 30 times and further correcting errors by calculating consensus[37].

We have studied the underlying algorithms of 107 read alignment tools that were designed for both short and long read sequencing technologies and were published in peer-reviewed articles from 1988 to 2020 (Table 1). Read alignment is a three-step procedure. First, it performs indexing with the aim of quickly locating genomic subsequences in the reference genome. This step includes building a large index database from a reference genome and/or the set of reads (Figure 1a,b). Second, it performs global positioning to determine the potential positions of each read in the reference genome. In this step, alignment algorithms use the prepared index to determine one or more possible regions of the reference genome that are likely to be similar to each read sequence (Figure 1c,d). Third, it performs pairwise alignment between the read and each of the corresponding regions of the reference genome to determine the exact number, location, and type of differences between the read and corresponding region (Figure 1e,f).

In our review, we define read alignment as a three step process, which includes indexing, global positioning, and pairwise alignment. In this case, pairwise alignment is considered to be performed between a read and a section of the reference determined by global positioning. Alternatively, the entire process can be viewed as local alignment with respect to the reference, and global alignment with respect to the read. In this formulation, the read is aligned end-to-end to the best substring in the reference and is expressed as semi-global alignment[38].

We have simplified pairwise alignment into overarching algorithm classifications like Smith-Waterman or Needleman-Wunsch, but tools that use dynamic programming can be classified into subcategories that are beyond the scope of this review. For example, read alignment algorithms can choose to be gapless (ignoring some variants), compute edit distance (the minimum number of edits needed to convert one string into the other), or use an affine gap penalty where variants are weighted differently based on their length. It is also worth noting that BWT-based tools do not use seeding in the traditional sense, and seed classification might be performed differently.

**Hashing is the most popular technique for indexing the reference genome**

The key goal of the indexing step is to facilitate quick and efficient querying over the whole reference genome sequence, producing a minimal memory footprint by storing the redundant subsequences of the reference genome only once[17,20,39]. Rapid advances in sequencing technologies have shaped the development of read alignment algorithms, and major changes in technology have rendered many tools obsolete. For example, some early methods[40–44] built the index database from the reads. Today's longer read lengths and increased throughput of sequencing technologies make such an approach infeasible for analyzing modern sequencing data. Modern alignment algorithms typically build the index database from the reference genome and then use the subsequences of the reads (known as seeds or qgrams) to query the index database (Figure 1a). In general, indexing the reference genome compared to the read set is a more practical and resource-frugal solution. Additionally, it allows reusing the constructed reference genome index across multiple samples.

We observe that the most popular indexing technique used by read alignment tools is hashing, which is used exclusively by 60.8% of our surveyed read aligner tools from various domains of biological research (Figure 2). Hashing is also the most popular individual indexing method for aligners that can handle DNA-Seq data, accounting for 68.3% of the surveyed read aligner tools. Hash table indexing was first used in 1988 by FASTA[45,46] and has since dominated the landscape of read alignment tools. Hashing was also the only dominant technique to be used until the BWT-FM index was introduced by Bowtie[47] (Figure 3a). Its popularity can be explained by the simplicity and ease of implementation when compared to other indexing techniques. Other advantages and limitations of hashing are outlined in Table 2. The hash table is a data structure that stores the content of some short regions of the genome (e.g., seeds) and their corresponding locations in the reference genome (Figure 1b). Such regions are also known as kmers or qgrams[48]. After the genomic seeds are produced, the alignment algorithm extracts the seeds from each read and uses them as a key to query the hash table index. The hash table returns a location list storing all occurrence locations of the read seed in the reference genome.

Indexing more seeds causes the size of the hash table to grow exponentially as the associated search space becomes larger (Table 2). For example, the human genome index when prepared with hashing is at least 1.5 times larger than the BWT-FM index (Supplementary Table 1). One solution is to consider fewer seeds to be indexed (e.g., non-overlapping consecutive seeds[49]). In general, the number of non-overlapping seeds selected by a read alignment algorithm determines the sensitivity (ability to report all *correct* mapping locations of a read) of the alignment algorithm, while the number of occurrences of all selected seeds determines the alignment speed[50]. To achieve a good balance between sensitivity and speed, read alignment methods attempt to select a large number of non-overlapping seeds while keeping the frequency of each seed relatively low[49–51].

**Alignment tools utilizing suffix-tree-based indexing are generally faster and more widely used**

The second most popular approach to indexing is the suffix-tree-based techniques, used exclusively by 36.5% of the surveyed read aligner tools (Figure 2) (Table 1). ERNE 2[52], LAMSA[53], and lordFAST[54] are categorized separately since they combine hashing with a suffix-tree-based technique. A suffix tree is a tree-like data structure where separate branches represent different suffixes of the genome; the shared prefix between the read and the genome is stored

only once and used by all the reads with that prefix. Every leaf node of the suffix tree stores all

occurrence locations of this unique suffix in the reference genome (Figure 1b). Unlike a hash

table, a suffix tree allows searching for both exact and inexact match seeds by walking through

the tree branches from the root to a leaf node, detouring as needed, following the query sequence

(Table 2). While some algorithms[55,56] specifically rely on creating suffix trees, the most

frequently chosen tools from this category use the Burrows-Wheeler Transform (BWT) and the

FM-index (hence called BWT-FM based tools) to mimic the suffix-tree traversal process while

generating a smaller memory footprint[57]. The FM-index compresses the BWT of the reference

genome, which is a representation of all permutations of the suffixes, and enables querying the

compressed BWT without decompression. The performance of the read aligners in this category

degrades as either the sequencing error rate increases or the genetic differences between the

subject and the reference genome are more likely to occur[58,59]. To allow mismatches, BWT-FM

aligners exhaustively traverse the data structure and match the seed to each possible path. This approach is impractical due to the rapid increase in the number of branches that need to be traversed as the number of mismatches increases. In order to reduce the tree traversal time and improve performance, many BWT-FM mappers perform a depth-first search (DFS) algorithm on the prefix tree and stop when the first hit is found within a certain user-defined threshold.

**The effect of read alignment algorithms on speed of alignment and computational**

**resources**

To measure the effect of read alignment algorithms on speed of alignment and computational resources, we have compared the running time and memory (RAM) required of eleven read alignment tools when applied to ten real WGS datasets (Figure 4a,b). We used tools available via the Bioconda package manager[60]. We ran these tools using their default parameters. We randomly selected ten WGS samples from the 1000 Genomes Project. We excluded tools specifically designed for RNA-Seq or BS-Seq. Details on how the tools were installed and ran are provided in Supplementary Note 2.

We found no significant difference in the runtime for BWT-FM tools and hashing based tools when adjusting for year of publication, chain of seeds, and type of pairwise alignment ($\beta = -0.11$; Likelihood ratio test (LRT p-value = 0.5) (Figure 4c, Supplementary Table 3,4). BWT-FM-based tools did require fewer computational resources when compared to hashing-based tools, adjusting for year of publication, chain of seeds, and type of pairwise alignment algorithm ($\beta = -1.51$; LRT p-value = $2.2 \times 10^{-3}$) (Figure 4d, Supplementary Table 5,6). The default suffix array implemented by LAST[51] requires increased running time and more computational resources when compared to BWT-FM-based tools ($\beta = 1.48$ and $1.27$ and LRT test p-value = $1.5 \times 10^{-15}$ and $<2 \times 10^{-16}$ for runtime and memory, respectively) (Figure 4c,d, Supplementary Table 3,4,5,6).

Despite the difference in performance driven by algorithms, we observed an overall improvement in computation time of read alignment with time ($\beta = -0.7$, s.e.=0.09; LRT test p-value=$3.7 \times 10^{-11}$) (Figure 4e, Supplementary Table 3,4) but no significant improvement of their memory requirements ($\beta = -0.21$, s.e.=0.24; LRT p-value=0.41) (Supplementary Figure 1, Supplementary Table 5,6). Usually, the index is created separately for each genome. Some methods incorporate multiple genomes into a single index graph[61–63], while other methods use a de Bruijn graph for hashing[52,62]. Although computing the genome index can take up to four hours, it usually needs to be computed only once and is often already precomputed for various species (Supplementary Figure 2). Updating the genome index can create a bottleneck in the analysis, especially for extremely large genome databases. Bloom-filter-based algorithms promise to provide an alternative way of indexing while preserving faster search times[64,65].

We surveyed 28 BWT-FM-based tools to compare the popularity of the read alignment algorithms using the number of times the introductory publication has been cited in other papers. Of those, three aligners have accumulated more than 1,000 citations per year since release, and 18% of the BWT-FM-based tools have been cited by at least 500 papers per year. In contrast, only two of the 63 hashing-based tools have more than 1,000 citations per year, but those two aligners (BLAST[66] and Gapped BLAST[67]) are, by far, the most popular with 2,726 and 3,143 citations per year, respectively (Figure 3b). Notably, tools cited more than 500 times per year were among the most effective both in terms of runtime and required computational resources (Supplementary Figure 3).

**Majority of the tools utilize fix length seeding to find the global position of the read in the reference genome**

The goal of the second step of read alignment is to find the global position of the read in the reference genome. This step is known as global positioning and uses the generated genome index to retrieve the locations (in the genome) of various seeds extracted from the sequencing reads (Figure 1c). The read alignment algorithm uses the determined seed locations to reduce the search space from the entire reference genome to only the neighborhood region of each seed location (Supplementary Note 1).

The number of possible locations of a seed in the reference genome is affected by two key factors: the seed length and the seed type. The estimated number of such locations is extremely large for short seeds and can reach tens of thousands for the human genome. The high frequency of short seeds is due to the repetitive nature of most genomes, which creates a high probability of finding the same short seed frequently in a long string of only four DNA letters. The large number of possible locations for short seeds imposes significant computational burden on read alignment algorithms[49,50].

Only a few read alignment algorithms examine all the seed locations reported in the location list[125]. Most of the read alignment algorithms apply heuristic devices to avoid examining all the locations of the seed in the reference genome (Figure 1d). First, read alignment tools can simply avoid examining seeds that occur more frequently than a user-defined threshold value[49,93,122,143]. Second, heuristics can examine all possible ways of dividing a query read into multiple seeds and select only the division that returns a set of infrequent seeds[153]. Third, instead of handling each seed hit independently, read alignment algorithms can rely on the coherence of locations for adjacent seeds over a region in the reference genome. This means that the read alignment algorithm requires adjacent seeds of the query read to also appear next to each other in the reference genome[49,126]. Finally, some other read alignment algorithms require each global position of a read in the reference genome to share a large number (more than a threshold) of short seeds with the read[113]. Most of the read alignment algorithms that follow these heuristics require choosing a large number of short seeds from each read.

Longer seed lengths can help reduce both the number of possible locations of a seed in the reference genome and the number of chosen seeds from each read. These benefits come at the cost of a possible reduction in alignment sensitivity, especially in cases where the mismatches between the read and the genome are located within the seed sequence. To enable increasing the seed length without reducing the alignment sensitivity, seeds can be generated as spaced seeds[48,69–72]. While a typical seed is a contiguous subsequence, a spaced seed contains in its sequence characters from a subsequence of the reference genome while ignoring the other characters of the same subsequence. Spaced seeds increase alignment sensitivity and enable hash tables to provide hits for both exact and inexact matches by ignoring certain bases of the seed. This approach was pioneered by PatternHunter[69–72] in 2002 and has been adopted by 14 tools. A majority of the tools using spaced seeds are designed for short read technologies (Table 1). Spaced seeds can also be used in long read alignment to tolerate high error rates[137]. Another approach to account for the error rate of sequencing technologies involves generating seeds as prefixes of the read sequence. Generating the prefixes of the reads—as opposed to generating the suffixes—allows the read alignment algorithm to tolerate an increased error rate towards the end of a read[154]. Other methods generate both suffix seeds and prefix seeds in order to tolerate large genetic variations[115].

The majority of the surveyed alignment algorithms use seeds of fixed length at run time. Some algorithms generate seeds of various lengths[107,130,155] in order to reduce the hit frequencies while tolerating mismatches. Varying the seed length or using different types of seed during the same run is often referred to as hybrid seeding[130] and was used by 20 of the 107 surveyed alignment algorithms. The first tool to use variable length seeds was GMAP[74]. Hybrid seeding with a hash-based index would require the creation of multiple hash tables of the same genome and would require extra computational resources. As a result, the vast majority of tools that use variable length seeds use a suffix tree indexing technique (BWT-FM or other).

Instead of choosing a large number of seeds from each read, read alignment algorithms can choose only a small number of seeds that are apart from each other. This approach also allows larger genetic variations and sequencing errors that are located between every two adjacent seeds[156]. Most read alignment algorithms that follow this approach try to limit the number of differences that are located at the gaps in order to avoid aligning a read to highly dissimilar regions in the reference genome. This approach can be performed using seed extension followed by seed chaining. First, after finding a matching seed shared between a read and the reference genome, the read alignment algorithm extends the matching seed in both directions until there are no more exact matches (such extended seeds are called maximal exact matches (MEMs)[157]). Second, the read alignment algorithm examines the gaps between every two adjacent extended seeds in the reference genome using a pairwise alignment algorithm[57,73] to construct a longer chain of these adjacent extended seeds [158]. The pairwise alignment can be performed end-to-end (e.g., global alignment) for two sequences of the same length[57,73], or by using a local alignment algorithm[40,121,159], where subsequences of the two given sequences are aligned. The two sequences can also be examined using a Hamming distance algorithm in cases where insertions or deletions are not allowed[99]. This seed chaining approach can also be applied to non-hashing-based read alignment algorithms, such as Bowtie2[160] and BWA-MEM[122]. We observe that 54 read alignment algorithms out of the 107 surveyed alignment algorithms use a seed chaining approach.

**Majority of the tools utilize Hamming distance and Smith-Waterman to determine similarity between the read and its global positions in the reference genome**

The goal of the last step of a read alignment algorithm is to determine regions of similarity between each read and the global positions of each read in the reference genome, which was determined in the previous step. These regions are potentially highly similar to the reads, but read alignment algorithms still need to determine the minimum number of differences between two genomic sequences, the nature of each difference, and the location of each difference in one of the two given sequences. Such information about the optimal location and the type of each edit is normally calculated using a verification algorithm (Figure 1f) that first verifies the similarity between the query read and the corresponding region in the reference genome. Verification algorithms can be categorized into algorithms based on dynamic programming (DP)[161] and non-DP based algorithms. The DP-based verification algorithms can be implemented as local alignment (e.g., Smith-Waterman[162]) or global alignment (e.g., Needleman-Wunsch[163]).

DP-based verification algorithms can also be implemented as semi-global alignment, where the entirety of one sequence is aligned to one of the ends of the other sequence[130,131,137].

The non-DP verification algorithms include Hamming distance[164] and the Rabin–Karp algorithm[165]. When one is interested in finding genetic substitutions, insertions, and deletions, DP-based algorithms are favored over non-DP algorithms. In general, the local alignment algorithm is preferred over global alignment when only a fraction of the read is expected to match with some regions of the reference genome due to, for example, large structural variations[90]. The Smith-Waterman[162] and Needleman-Wunsch[163] alignment algorithms were both first used by FASTA[45,46] in 1988, which we categorize as "Multiple Methods" (Figure 3c). Smith-Waterman remains the most popular algorithm and is used by 28.3% of our surveyed tools (Figure 2). Needleman-Wunsch, in contrast, has only been used by 16.2% of our surveyed tools (Figure 2). However, if we include the tools which allow for multiple methods, Smith-Waterman represents 38.3% and Needleman-Wunsh represents 26.2% of alignment algorithms used. This trend is due to the fact that 12 of the 13 tools classified as "Multiple Methods" use or allow both Smith-Waterman and Needleman-Wunsch. Non-DP verification using Hamming distance[164] has been the second most popular single technique since used for the first time by RMAP[42] in 2008 (Figure 3c). There is no significant correlation between the indexing technique used and the pairwise alignment algorithm chosen. Most major indexing techniques are used in conjunction with most pairwise alignments. However, BWT-FM-based aligners do comprise the largest percentage of tools that allow multiple pairwise alignment methods (Figure 2).

As the number of differences between two sequences is not necessarily equivalent to the sum of the number of differences between the sub-sequences of these sequences, it is necessary to perform verification for the entire read sequence and the corresponding region in the reference sequence[166]. Existing DP-based algorithms can be inefficient as they require quadratic time and space complexity. Despite more than three decades of attempts to improve their algorithmic implementation, the fastest known edit distance computation algorithm is still nearly quadratic[167]. Some of the read alignment algorithms use DP only for seed chaining, which provides suboptimal alignment calculation[35,73]. This approach is called sparse DP and is used in C4[73], conLSH[152], and LAMSA[53].

An alternative way to accelerate the alignment algorithms is by reducing the maximum number of differences that can be detected by the verification algorithm, which reduces the search space of the DP algorithm and shortens the computation time[168–170]. Modern read alignment algorithms (e.g., Hobbes[153], Hobbes2[131], Bitmapper[171], mrFAST[125], RazerS[89]) exploit this observation to develop heuristics that quickly decide whether or not the computationally expensive DP calculation is needed—if not, significant time is saved by avoiding DP calculation. Such heuristics are called *pre-alignment filters*[172–175], and they approximate the total number of differences between two sequences to determine if this count is greater than a threshold (Figure 1e). If so, these heuristics decide that the verification calculation is not needed due to high dissimilarity between the two sequences. Verification algorithms can also be accelerated using specialized or general-purpose hardware accelerators such as multi-core processors[176,177,178].

We found that tools which use the Needleman-Wunsch[163] algorithm are faster than tools which use other algorithms ($\beta$ = 1.37, 1.22, and 0.78 and Wald test p-values $9.3x10^{-7}$, $1.8x10^{-10}$, and $1.3x10^{-4}$ for Hamming distance, non-DP heuristics, and SW algorithms, respectively) (Figure 4f, Supplementary Table 3), adjusting for publication year, seed chaining, and indexing method. Despite the overall longer runtime of Hamming-distance-based methods, the latest hashing based tools (e.g. HISAT2[150]) provide a comparable running time with the fastest Needleman-Wunsch-based tools. We also found significant differences in the amount of computational resources required by read alignment tools using different pairwise alignment algorithms after adjusting for publication year, type of seed, and indexing method (LRT; p-value = 0.04) (Supplementary Figure 4, Supplementary Table 6). Notably, the algorithms with the smallest computational footprints use various types of pairwise alignment algorithms.

After examining all global positions of a given read in the reference genome, the read can still align to multiple global positions in the reference genome equally well[29]. Such reads are known as multi-mapped reads. Read alignment algorithms employ several strategies to deal with multi-mapped reads. Some read alignment algorithms simply discard these multi-mapped reads (e.g., SNAP[105]). Other algorithms perform additional computation to make a deterministic choice for the matching regions (e.g., RazerS[113]). Another popular technique is to select one of the locations randomly (e.g., BWA-MEM[122] and Stampy[106]), in which case re-running the same tool may produce non-reproducible results in the form of different mapping outputs[29].

Read alignment algorithms tend to ignore information about genome coverage. Few methods attempt to employ a coverage profile in order to inform the decision of where to align the read[126].

Few read alignment algorithms attempt to remove redundancies in the read set before performing read alignment in order to improve the overall alignment time by avoiding re-alignment of the same read sequence[39,43].

**Influence of long read technologies on the development of novel read alignment algorithm**

Alignment of the long reads produced by modern long-read technologies[16,37,156] provides a unique possibility to discover previously-undetectable structural variants[16,179,180]. Long reads also improve the construction of an accurate hybrid *de novo* assembly[16,181], in cases where long and short reads are suffix-prefix overlapped, or in cases where reads are aligned using pairwise alignment algorithms, to construct an entire assembly graph. This is helpful when a reference genome is either unavailable[182,183] or is complex and contains large repetitive genomic regions[184].

Existing long read alignment algorithms still follow the three-step based approach of short read alignment. Some long read alignment tools even divide every long read into short segments (e.g., 250 bp), align each short segment individually, and determine the mapping locations of each long read based on the adjacent mapping locations of these short segments[142,145]. Some long read alignment tools use hash-based indexing[132,140,185], while others use BWT-FM indexing[83,122,186]. The major challenge with the long read alignment algorithms are dealing with large sequencing errors and a significantly large number of short seeds extracted from each long or ultra-long read[187]. Thus, the most recently developed long read alignment algorithms require heuristically extracting fewer seeds per read length when compared to those extracted from short reads.

Instead of creating a hash table for the full set of seeds, recent long read alignment algorithms find the minimum representative set of seeds from a group of adjacent seeds within a genomic region. These representative seeds are called minimizers[188,189] and can also be used to compress genomic data[190] or taxonomically profile metagenomic samples[191]. One way to generate a minimizer is to extract all overlapping seeds from a short genomic subsequence, then lexicographically sort them, and choose a single seed as a minimizer. This process is repeated for all short genomic subsequences of the long read. Long read alignment algorithms[139,143,192] that use hashed minimizers as an indexing technique provide a faster alignment process compared to other algorithms that use conventional seeding or BWT-FM. They also provide a significantly faster (>10x) indexing time (Supplementary Table 1). However, their accuracy degrades with the use of short reads as they process fewer number of seeds per short read[143].

**Box 1. Advantages and limitations of short versus long read alignment algorithms**

- Error rate. Error rate of modern short read sequencing technologies is smaller than that of modern long read technologies.

- Genome coverage. Throughput (i.e., the number of reads) of modern short read sequencing technologies is higher than that of modern long read technologies.

- Global position. Determine a global position of the read by identifying the starting position or positions of the reads in the reference genome. This step is ambiguous with short reads, as the repetitive structure of the human genome causes such reads to align to multiple locations of the genome. In contrast, long reads are usually longer than the majority of repeat regions and are aligned to a single location in the genome.

- Local pairwise alignment. After determining the global position of each read, the algorithmsmap all bases of the read to the reference segments, located at these global positions, in order to account for indels. Due to the smaller error rate of short read technologies, it is usually easier to perform local alignment on short reads than on long ones.

- Genomic variants. Single Nucleotide Polymorphisms (SNPs) are easy to detect using short reads when compared to long reads due to the lower error rate and higher coverage of short read sequencing technologies. Structural variants (SVs) are easy to detect with long reads, which span the entire SV region. Current long-read-based tools[193] are able to detect deletions and insertions with high precision. The sparse coverage of long reads may lower the sensitivity of detection.

 -- End box

## Read alignment across various domains of biological research

We discuss the challenges and the features of these algorithms that are specific to the various domains of modern biological research. Often the domain-specific alignment problem can be solved by creating a novel tool from scratch or wrapping the existing algorithms into a domain-specific alignment tool (Supplementary Figure 5 and 6). Additionally, longer reads make the read alignment problem similar across areas of biological research. For example, tools recently designed to align long reads can handle both DNA and RNA-Seq reads[148].

## RNA-Seq alignment

RNA-Sequencing is a technique used to investigate transcriptomics by generating millions of reads from a collection of human alternative spliced isoforms transcripts, referred to as a transcriptome[194]. RNA-Seq has been widely used for gene expression analysis as well as splicing analysis[194,195,14]. However, the alignment of RNA-Sequencing reads needs to overcome additional challenges when mapping the reads originating from human transcriptome onto the reference genomes. Those challenges arise due to differences between the human transcriptome and the human genome; these differences define a subset of alignment problems known as *spliced alignment*. Spliced alignment requires that the researcher take into account reads spanning over large gaps caused by spliced out introns[194]. Reads spanning only a few bases across the junctions can be easily aligned to an adjacent intron or aligned in a wrong location, making the accurate alignment more difficult[14,194].

Several spliced alignment tools have been developed to address this issue and align RNA-Seq reads in a splicing-aware manner (Table 1 and Figure 1c). Hashing is the most popular technique among RNA-Seq aligners (Supplementary Figure 7). This is even more evident if we remove the RNA-Seq aligners that are wrappers of existing DNA-Seq alignment methods (Supplementary Figure 5). Over 60% of the RNA-Seq aligners which are wrappers of existing DNA-Seq alignment methods use Bowtie or Bowtie2 (Supplementary Figure 5). When considering only stand alone RNA-Seq aligners, the number of aligners using hashing more than doubles the number of aligners using an FM-index (Supplementary Figure 8).

The most popular tool based on the number of citations was TopHat2[128] (Table 1). TopHat2 uses Bowtie2 to align reads that may span more than one exon by splitting the reads into smaller segments and stitching the segments together to form a whole read alignment. The stitched read alignment spans a splicing junction on the human genome. This method allows identification of the splicing junction without transcriptome alignment. A more recent tool, HISAT2, uses a hierarchical indexing algorithm that leverages the Burrows-Wheeler Transform and Ferragina-Manzini index to align parts of reads and extend the alignment[63]. Another popular method, RNA-Seq aligner—called STAR—utilizes suffix arrays to identify a maximal mappable prefix, which is used as seeds or anchors, and stitch together the seeds that aligned within the same genomic window[127]. Although those tools can detect splicing junctions within their algorithm, it is possible to supply known gene annotation to increase the accuracy of a spliced alignment. The alignment accuracy, measured by correct read placement, can be increased 5-10% by supplying known gene annotations[14,194]. HISAT2 and STAR are able to align the reads accurately with or without a splicing junction[14]. Furthermore, the discovery and quantification of novel splicing junctions can be significantly improved using two passes in STAR, which generates a list of possible junctions in the first pass and identifies aligning reads leveraging the junctions in the second pass[196]. While spliced alignment can provide an important splicing junction information, those tools require intensive computational resources[14].

To align RNA-Seq reads onto the transcriptome reference instead of the genome reference, regular DNA aligners are typically used. Mapping to the transcriptome is usually performed to estimate expression levels of genes and alternatively spliced isoforms by assigning reads to genes and alternatively spliced isoforms[127,197]. Since many alternatively spliced isoforms share exons which are usually longer than the short reads, probabilistic models are used as it is impossible to uniquely assign reads to the isoform transcripts[198].

Alternatively, one can avoid computationally expensive alignment and perform pseudo-alignment. Pseudo-alignment is a type of k-mer index that looks up k-compatibility class for each k-mer in the read and the index, enabling the user to quantify gene expressions in an alignment-free manner[199]. In contrast to regular base-level alignment algorithms, pseudo-alignment algorithms[199,200] are unable to provide the precise alignment position of the read in the genome. Instead, pseudo-alignment algorithms assign the reads to a corresponding gene and/or alternatively spliced isoform. Usually, such information is sufficient to accurately estimate gene expression levels of the sample[201].

**Metagenomic alignment**

Metagenomics is a technique used to investigate the genetic material in human or environmental microbial samples by generating millions of reads from the microbiome—a complex microbial community residing in the sample. Metagenomic data often contains an increased number of reads required to be aligned against more than hundreds of thousands of microbial genome. For example, as of July 2018, the total number of nucleotides in NCBI's collection of bacterial genomes measures over 204 times the number of nucleotides present in the Genome Reference Consortium Human Build 38 (Supplementary Note 3). The increased number of reads and the size of reference databases pose unique challenges to existing alignment algorithms when applied to metagenomics studies.

In targeted gene sequencing studies, such as those that sequence portions of the 16S ribosomal RNA of prokaryotes or internally transcribed spacers (ITS) of eukaryotes, a number of task-specific aligners are utilized to identify the origin of candidate reads or to perform homology searches. For example, Infernal[202] utilizes profile hidden Markov models to perform alignment based on RNA secondary structure information. Multiple sequence aligners are also utilized in metagenomic analysis pipelines such as QIIME[203], Mothur[204], and Megan[204,205]. For example, NAST[204–206] and PyNAST[207] use 7-mer seeds and a BLAST alignment that is then further refined using a bidirectional search to handle indels. Similarly, MUSCLE[207,208] uses an initial distance estimation based on k-mers and proceeds through a progressively constructed hierarchical guide tree while optimizing a log expectation for multiple sequence alignment[208].

For untargeted whole genome shotgun (WGS) metagenomic studies, the task of identifying the genomic or taxonomic origin of sequencing reads (referred to as "fragment recruitment" or "taxonomic read binning") is even more difficult, Individual reads can originate from multiple organisms due to shared homology or horizontal gene transfer and reads may originate from previously unsequenced organisms. This has sparked the development of a variety of tools[209] which aim to identify the presence and relative abundance of taxa or organisms present in a metagenomic sample via a reference-free and/or alignment-free fashion (referred to as "taxonomic profiling"). Similar in spirit to RNA-Seq alignment, these tools avoid computationally expensive base-level alignment and perform pseudo-alignment or multiple types of k-mer matching to detect the presence of organisms in a metagenomic sample[191,210,211], as well as use minimizers to reduce computational time[191].

Other approaches handle growing reference database sizes by aligning reads onto a reduced reference database, sometimes composed of marker microbial genes that are present in a specific taxa. Reads mapping to those genes can be used to determine the presence of specific taxa in a sample[212]. Such tools typically use existing DNA alignment algorithms.;or example, MetaPhlAn[212] uses the Bowtie2 aligner.

Even with the development of these new metagenomic tools, existing read alignment tools (e.g., MOSAIK, SOAP, and BWA) are still used for fragment recruitment purposes[213]. However, use of existing read alignment tools for metagenomics carries a significant computational burden and is identified as the main bottleneck in the analysis of such data. This major limitation suggests the need for development of alignment tools capable of handling the increased number of reads and reference genomes seen in such studies [214].

Metagenomics studies are also capable of functional annotation of microbiome samples by aligning the reads to genes, gene families, protein families, or metabolic pathways. Protein alignment is beyond the scope of this manuscript, but many of the algorithmic approaches previously discussed are utilized for functional annotation[213,215]. For example, RAPSearch2[213,215] uses a collision free hash table based on amino acid 6-mers. The protein aligner DIAMOND[216] utilizes a spaced-seed-and-extend approach based on a reduced alphabet and a unique indexing of both reference and query sequences. Indexing of both the reference *and* the query reads provides multiple orders of magnitude in speed improvements over older tools (such as BLASTX) at the cost of increased memory usage. Recently, MMseqs2[214] utilizes consecutive, similar k-mer matches to further improve the speed of protein alignment.

**Viral quasispecies alignment**

RNA viruses such as Human Immunodeficiency virus (HIV) are highly mutable, with the mutation rates being as high as $10^{-4}$ per base per cell[217] allowing such viruses to form highly heterogeneous populations of closely related genomic variants commonly referred to as quasispecies[218]. Rare genomic variants, which are a few mutations away from the major strain, are often responsible for immune escape, drug resistance, viral transmission, and increase of virulence and infectivity of the viruses[219-220]. Massively parallel sequencing techniques allow for sampling of intra-host viral populations at high depth and provide the ability to profile the full spectra of viral quasispecies, including rare variants.

Similar to other domains, accurate read alignment is essential for assembling viral genomic variants including the rare ones. Aligning reads that originated from heterogeneous populations of closely related genomic variants to the reference viral genome give rise to unique challenges for existing read alignment algorithms. For example, read alignment methods should be extremely sensitive to small genomic variations while being robust to artificial variations introduced by sequencing technologies. At the same time, the genetic difference between viral quasispecies of different hosts is usually substantial (unless they originated from the same viral outbreak or transmission cluster), which makes application of predefined libraries of reference sequences for viral read alignment problematic or even impossible.

Currently available tools for viral haplotyping (e.g., CliqueSNV[221], 2SNV[222], PredictHaplo[223], aBayesQR[224], QuasiRecomb[225], ShotMCF[226], SHORAH[227]) and variant calling (e.g., V-Phaser 2[228], VirVarSeq[229], MinVar[230]) frequently rely on existing independent alignment tools. While viral samples contain several distinct haplotypes, the read alignment tools such as BWA[59] and BowTie[231] can only map reads to a single reference sequence. Since certain haplotypes may be further or closer to the reference sequence, the reads emitted by such haplotypes may have quite different mapping quality. Some tools re-align reads to the consensus sequence instead of keeping the original alignment to the reference. Nevertheless, even alignment to the perfect reference or consensus sequence can reject perfectly valid short reads because of multiple mismatches. Rejection of such reads may cause loss of rare haplotypes and mutations.

Systematic sequencing errors (such as homopolymer errors) frequently cause alignment errors. Although the sequencing error rate, both systematic and random, is comparatively low, such errors can be more frequent than the rarest variants. The alignment errors caused by sequencing errors may cause drastic sensitivity and reduction in specificity of haplotyping and variant calling methods. Supplementary Figure 9 provides an example of an alignment of Illumina reads produced by Influenza A virus quasispecies. In this case, true viral haplotypes contain deletions that are inconsistently aligned to the reference. Such inconsistencies cause erroneous single-position shifts in the alignment, which in turn results in discovery of false positive single nucleotide variations.

As a result, many viral haplotyping and variant calling methods utilize customized pre-processing algorithms prior to the read alignment step. The most popular strategies include separation of reads into clusters and generation of consensus sequences for the clusters to be used as local references [232] and k-mer-based preprocessing of reads to eliminate the most obvious sequencing errors[233].

**Aligning bisulfite-converted sequencing reads**

Bisulfite-converted sequencing is a technique used to sequence methylated fragments[234,235]. During sequencing, most of the cytosines (C) in the reads become thymines (T). Since every sequenced T could either be a genuine genomic T or a converted C, special techniques are used to map those reads[236]. Some tools substitute all C in reads with wildcard bases, which can be aligned to C or T in the reference genome[72,81], while other tools substitute all C by T in all reads and reference and work with a three-letter alphabet aligning to a C-to-T-converted genome[103,120]. Unlike RNA-Seq aligners, FM-index was the most popular technique among BS-Seq aligners (Supplementary Figure 10). One third of the surveyed BS-Seq aligners were wrappers of existing DNA-Seq alignment methods (Supplementary Figure 6), with all three of those wrapping Bowtie or Bowtie2 (Supplementary Figure 6). As a result, when considering only stand-alone BS-Seq aligners, the numbers of aligners using each indexing algorithm become extremely similar (Supplementary Figure 11).

**Other domains**

Other domains requiring specialized alignment include B and T cell receptor repertoire analysis. The repertoire data is generated using targeter repertoire sequencing protocols, known as BCR-or TCR-Seq. For example, tools designed to align reads to the V(D)J genes use combinations of fast alignment algorithms and more sensitive modified Smith–Waterman and Needleman–Wunsch algorithms[191,237,238].

**Discrepancies between the reads and the reference may reveal the historical errors in the reference assembly**

Genome sequencing datasets, especially those generated with long reads, provide a unique perspective to reveal errors in the reference assemblies (e.g., human reference genome) based on the discrepancies between the reads and the reference sequence. References and reads (e.g., resequencing data) are often produced using different technologies, and there are usually disagreements between references and reads that produce mapping errors. Similarly, some of these errors also come from the errors in the reads used for assembly, collapsed/merged duplications/repeats, and heterozygosity. For example, a study for structural variation discovery led to identification of incorrectly inverted segments in the reference genome[239]. Similarly, Dennis et al.[240] characterized a duplicated gene that was not represented accurately because it collapsed in the reference genome. Therefore, using the most recent version of a reference genome is always the best practice, as demonstrated by an analysis of the latest version of the human genome[240,241].

Structural errors in the reference genomes can be found and corrected by using various orthogonal technologies such as mate-pair and paired-end sequencing[242,243], optical mapping[244], and linked-read sequencing[245]. Smaller-scale errors (i.e., substitutions and indels) can also be corrected using assembly polishing tools such as Pilon, which employs short read sequencing data[246]. However, long reads are more powerful in detecting and correcting errors due to the fact that they can span most common repeat elements. Long read based assembly polishers include Quiver[247] that uses Pacific Biosciences data, Nanopolish[248] that uses Nanopore sequencing, and Apollo[249] that can use read sets from any sequencing technology to polish large genomes. Additionally, more modern long read genome assemblers, such as Canu[250], include built-in assembly polishing tools.

**Discussion**

Read alignment is one of the most commonly-used steps in most bioinformatics analyses due to the inability of sequencing machines to read the complete genome as a single string. We are witnessing an exciting time in bioinformatics during which rapid advances in sequencing technologies shape the landscape of modern read alignment algorithms. Changes in sequencing technologies can render some read alignment algorithms irrelevant—yet provide context for the development of new tools that promise to address the sequencing inaccuracy of modern sequencing machines. The development of alignment algorithms is shaped not only by the characteristics of sequencing technologies but also by the specific characteristics of the application domain. Often different biological questions can be answered using similar bioinformatics algorithms. For example BLAT[35,251], a tool which was originally designed to map EST and Sanger reads, is now used to map the assembled contigs to the reference genome[251]. Specific features of various domains of biological research, including whole transcriptome, adaptive immune repertoire, and human microbiome studies, confront the developer with a choice of developing a novel algorithm from scratch or adjusting existing algorithms.

In general, the read alignment problem is extremely challenging due to the large size of analyzed datasets and numerous technological limitations of modern sequencing platforms. A modern read aligner should not only be able to maintain a good balance between speed and memory usage, but also be able to preserve small and large genetic variations. It should be capable of tackling numerous biotechnological limitations and changes, ultimately inducing rapid evolution of sequencing technologies such as constant growth of read length and changes in error rates. In general, determining an accurate global position of the read in the reference genome provides no guarantee that accurate local pairwise alignment can be found. This is especially challenging for the error-prone long reads, where determining accurate global position of the read in the reference genome is usually easy, but local pairwise alignment represents a substantial challenge due to a high error rate.

Looking forward, with the development of new sequencing machines (e.g., Illumina NextSeq 2000) that perform both sequencing and real-time read alignment on the same machine, we are also going to witness new specialized hardware architectures for read alignment that can cope with the growing throughput of these sequencing machines. This approach has two benefits. First, it can hide the complexity and details of the underlying hardware from users who are not necessarily fluent in handling complex hardware[252]. Second, it helps to reduce the total analysis time[252,253], by starting read alignment while still sequencing.

Looking into the late future, even if accurately sequencing the entire genome as a single string might be possible, we believe that most of the techniques involved in current read alignment algorithms will continue to remain a crucial component in analyzing and comparing the sequencing data. For example, we still need to quickly and efficiently compare a complete genome of an unknown donor to a set of complete genomes for metagenomic profiling. In this case, comparing genomes of the same length may not provide accurate profiling due to possible large genetic variations. Similarly, we still need to know the length and the location of possible genetic variations between individuals' genomes in a population. This requires developing efficient algorithmic methods to quickly find these variations without the need to perform DP-based pairwise alignment between the complete genomes.

We hope the review in this work on fundamentals and recent research not only provides an understanding of the basic concepts of read alignment, its limitation, and how they are mitigated, but also helps inform its future directions in read alignment development. We believe the future is bright for read alignment algorithms, and we hope that the many examples of read alignment algorithms presented in this work inspire researchers and developers to enhance future read alignment.

**Availability of Materials**

All data and code required to produce the figures contained within this text are freely available on GitHub: https://github.com/Mangul-Lab-USC/review.technology.dictates.algorithms.

## Acknowledgments

We thank the authors of the tools surveyed in this work for providing helpful feedback and verifying the information related to their tool. We also thank Martin Frith (University of Tokyo), Heng Li (Harvard University), Cenk Sahinalp (National Cancer Institute), and Steven Salzberg (Johns Hopkins University) for their valuable feedback and discussion. B.D.S. is supported by NIH/NHLBI K08HL128867, P.S. is supported by NIH 1R01EB025022, P.I.B. and S.K. are supported by the Molecular Basis of Disease (MBD), and O.M. is supported by Intel, VMware, and NIH HG006004. The authors acknowledge the Computational Genomics Summer Institute, funded by NIH GM112625, which fostered international collaboration among the groups involved in this project.

## Author Contributions

S.M. led the project, B.B. performed statistical analysis, and J.R. and M.A. produced the figures. H.S., J.R., K.T., and M.A. compiled Table 1. J.R., P.I.B., and V.X. created scripts for running and evaluating software tools. A.Z., B.B., B.D.S., C.A., D.K., H.S., H.T.Y., J.R., M.A., O.M., P.S., S.K., and S.M. wrote, reviewed, and edited the manuscript.

## Competing Interests

{Figures and Figure Legends}

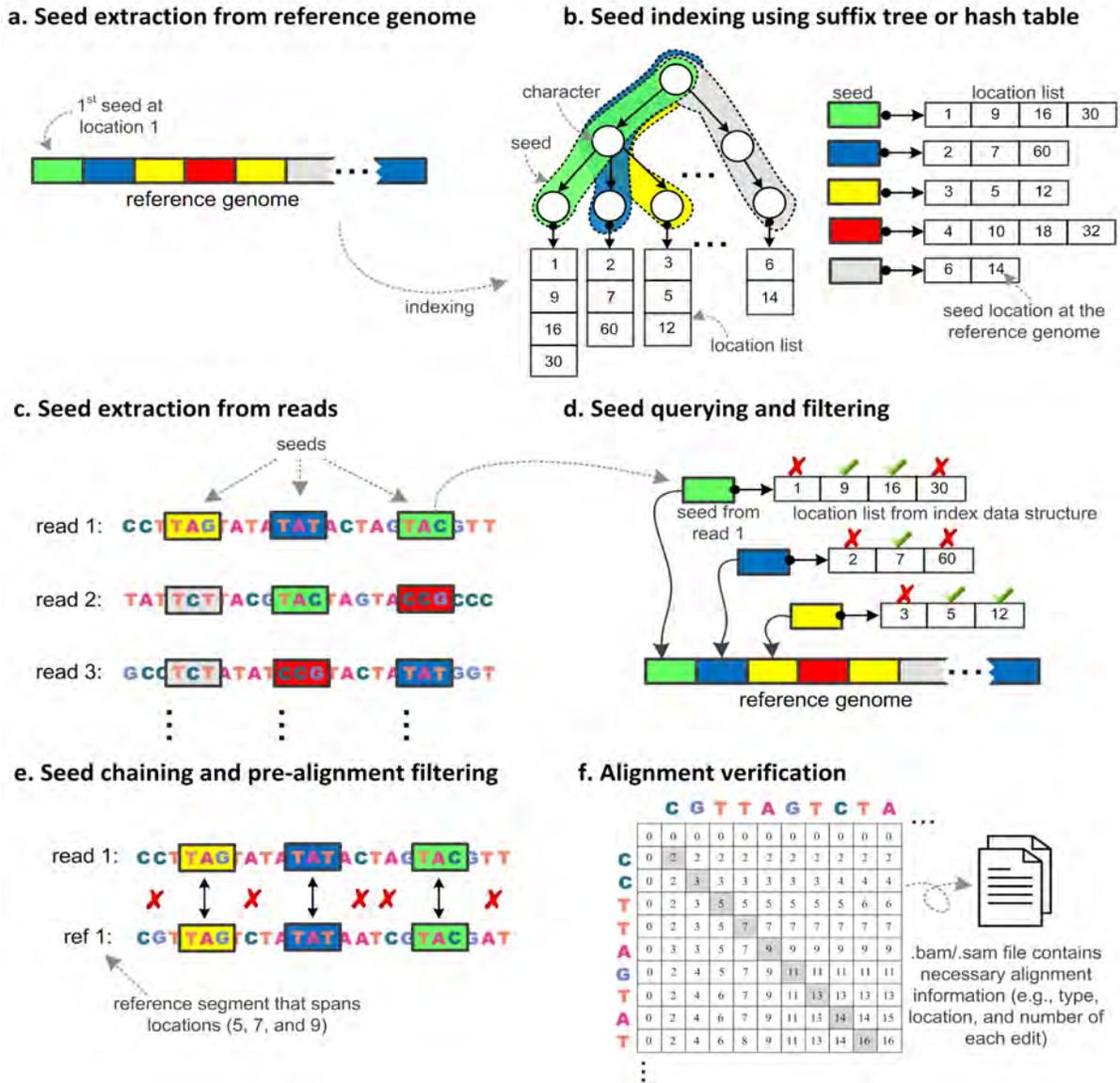

**Figure 1. Overview of a read alignment algorithm.** (a) The seeds from the reference genome sequence are extracted. (b) Each extracted seed and all its occurrence locations in the reference genome are stored using data structure of choice (suffix tree and hash table are presented as an example). Common prefixes of the seeds are stored once in the branches of the suffix tree, while hash table stores each seed individually. (c) The seeds from each read sequence are extracted. (d) The occurrences of each extracted seed in the reference genome are determined by querying the index database. In this example, the three seeds from the first read appear adjacent at locations 5,

7, and 9 in the reference genome. Two of the same seeds appear also adjacent at another two locations (12 and 16). Other non-adjacent locations are filtered out (marked with X) as they may not span a good match with the first read. (e) The adjacent seeds are linked together to form a longer chain of seeds by examining the mismatches between the gaps. Pre-alignment filters can also be applied to quickly decide whether or not the computationally expensive DP calculation is needed. (f) Once the pre-alignment filter accepts the alignment between a read and a region in the reference genome then DP-based (or non-DP based) verification algorithms are used to generate the alignment file (in BAM or SAM formats), which contains alignment information such as the exact number of differences, location of each difference, and their type.

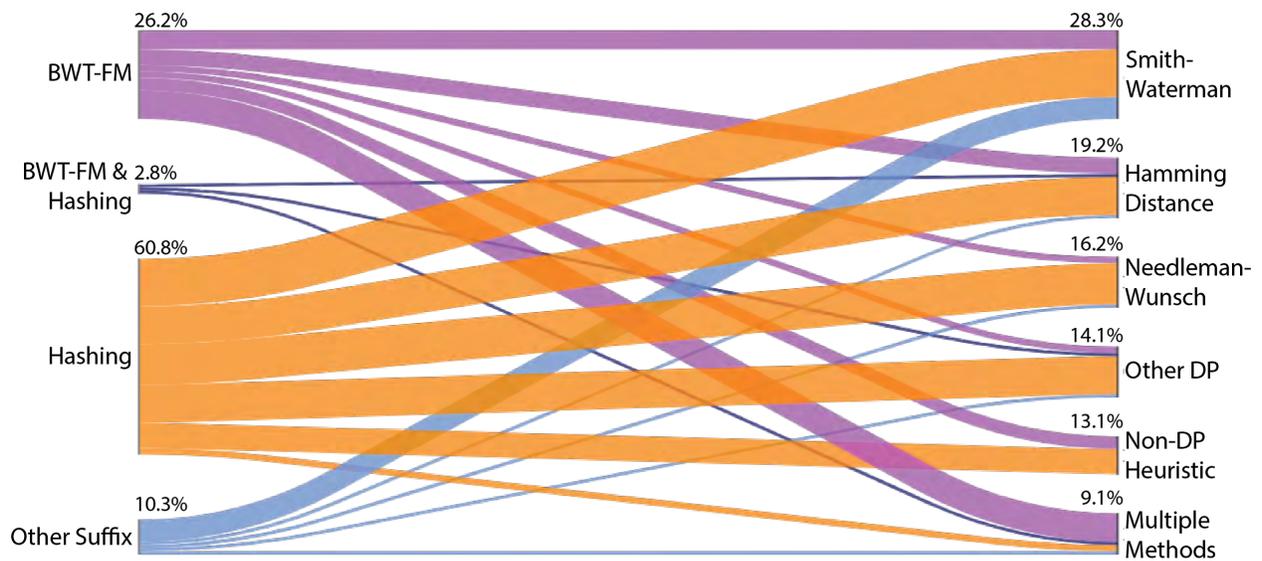

**Figure 2. Combination of algorithms utilized by read alignment tools.** Sankey plot displaying the flow of surveyed tools using each indexing technique and pairwise alignment. For every indexing technique, the percentage of surveyed tools using the algorithm is displayed (BWT-FM: 26.2%, BWT-FM and Hashing: 2.8%, Hashing: 60.8%, Other Suffix: 10.3%). For every pairwise alignment technique, the percentage of surveyed tools using the algorithm is displayed (Smith-Waterman:28.3%, Hamming Distance: 19.2%, Needleman-Wunsch: 16.2%, Other DP: 14.1%, Non-DP Heuristic: 13.1%, Multiple Methods 9.1%).

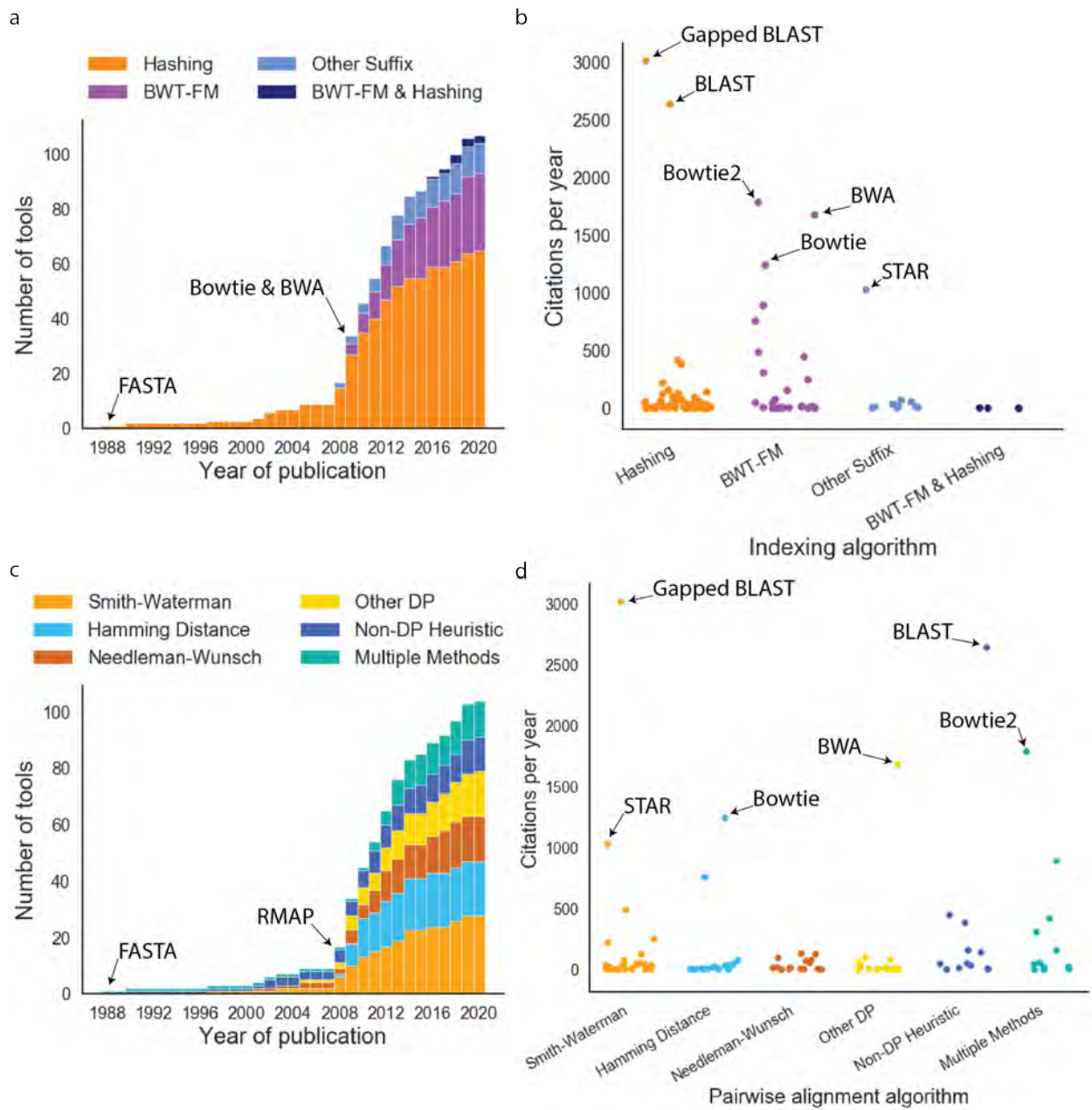

**Figure 3. Landscape of read alignment algorithms published from 1988 to 2020.** (a) Histogram showing the cumulation of surveyed tools over time colored by the algorithm used for genome indexing. The first published aligner, FASTA, is labeled as well as the point at which Bowtie and BWA were introduced and changed the landscape of aligners. (b) The popularities of all surveyed aligners, judged by citations per year since initial release. Tools are grouped by the algorithm used for genome indexing. The six overall most popular aligners are labeled. (c) Histogram showing the cumulation of surveyed tools over time colored by the algorithm used for pairwise alignment. The two aligners credited to have been the first to use the three most popular algorithms (FASTA: Smith-Waterman and Needleman-Wunsch, RMAP: Hamming Distance) are labeled. (d) The popularity of each surveyed aligner, judged by citations per year since initial release. Tools are grouped by the algorithm used for pairwise alignment. The six overall most popular aligners are labeled.

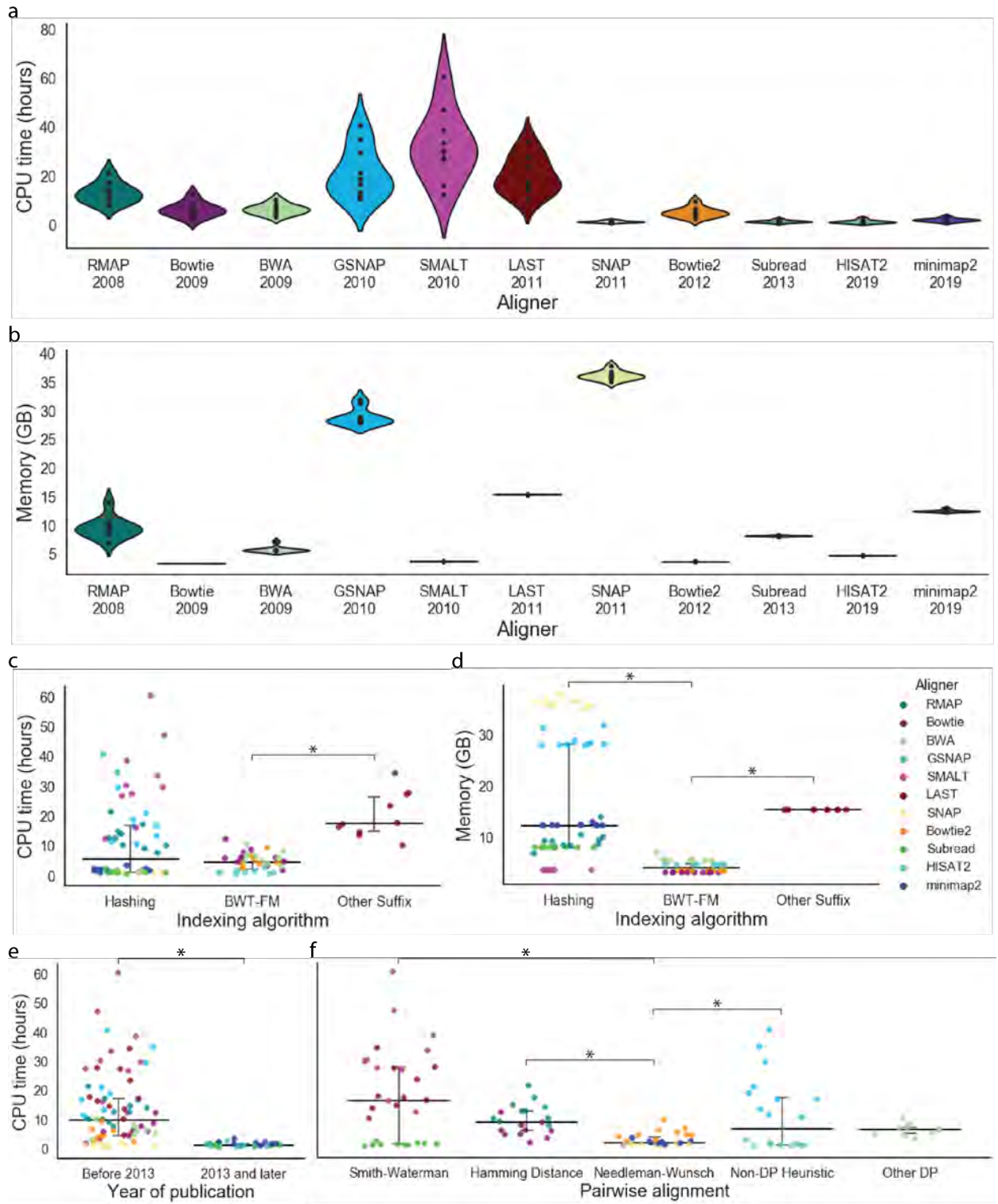

**Figure 4. The effect of read alignment algorithms on speed of alignment and computational resources.** Results of the benchmarking performed on 11 surveyed DNA read alignment tools

that can be installed through bioconda (RMAP, Bowtie, BWA, GSNAP, SMALT, LAST, SNAP, Bowtie2, Subread, HISAT2, and minimap2) additionally noted in Supplementary Table 2 and Supplementary Note 2. Each tool's CPU time and RAM required was recorded for 10 different WGS samples from the 1000 Genomes Project. (a-b) Violin plots showing the relative performance ((a) CPU time and (b) RAM) of the benchmarked aligners. Aligners are ordered by year of release. (c-d) The relative performance ((c) CPU time and (d) RAM) of the benchmarked aligners grouped by the algorithm used for genome indexing and colored by individual aligners (BWT-FM CPU time vs. Suffix Array CPU time: LRT, $p$-value = $1.5 \times 10^{-15}$, Hashing memory vs. BWT-FM memory: LRT, $p$-value = $2.2 \times 10^{-3}$, BWT-FM memory vs. Suffix Array memory: LRT, $p$-value < $2 \times 10^{-16}$). (e) The relative performance (CPU time) of the benchmarked aligners grouped by whether the tool was released before or after long read technology was introduced (2013) and colored by individual aligners (LRT, $p$-value = $3.7 \times 10^{-11}$). (f) The relative performance (CPU time) of the benchmarked aligners grouped by the algorithm used for pairwise alignment and colored by individual aligners (Needleman-Wunsch CPU time vs. Smith-Waterman CPU time: Wald, $p$-value = $1.3 \times 10^{-4}$, Needleman-Wunsch CPU time vs. Hamming Distance CPU time: Wald, $p$-value = $9.3 \times 10^{-7}$, Needleman-Wunsch CPU time vs. Non-DP Heuristic CPU time: Wald, $p$-value = $1.8 \times 10^{-10}$).

**Table 1. Summary of algorithms and features of the examined read alignment methods.** We surveyed 107 alignment tools published from 1988 to 2020 (indicated in column "Year of publication"). The table is sorted by year of publication, and then grouped according to the area(s) of application (indicated in column "Application") within each year. In column "Indexing", we document the algorithms used to index the genome (the first step in read alignment). In column "Global Positioning", we document the algorithms used to determine a global position of the read in the reference genome (the second step). In column "Pairwise alignment", we document the algorithm used to determine the similarity between the read and the corresponding region of the reference genome (the last step). SW, NW, HD, DP stand for Smith-Waterman algorithm, Needleman-Wunsch algorithm, Hamming distance, and dynamic programming, respectively. In column "Wrapper", we document the read alignment algorithms that are built on top of other read alignment tools. Finally, we report the maximum read length tested in the corresponding paper in column "Max. Read Length Tested in the Paper (bp)". The tested read length in each paper is not necessarily the maximum read length that each tool can handle.

| Aligner | URL | | | Indexing | Global Positioning | Pairwise alignment | Wr | Max. Read |
|---------|-----|---|---|----------|--------------------|--------------------|-----|-----------|

| | | Year of publication | Application | | Fix length seed | Spaced seed | Seed chaining | | apper | Length Tested in the Paper (bp) |
|---|---|---|---|---|---|---|---|---|---|---|
| FASTA[46] | https://fasta.bioch.virginia.edu/fasta_www2/fasta_list2.shtml | 1988 | DNA | hashing | Y | N | Y | SW and NW | N | 1500 |
| BLAST[66] | https://blast.ncbi.nlm.nih.gov/Blast.cgi | 1990 | DNA | hashing | Y | N | Y | Non-DP Heuristic | N | 73360 |
| Gapped BLAST[67] | https://blast.ncbi.nlm.nih.gov/Blast.cgi | 1997 | DNA | hashing | Y | N | Y | SW | N | 246 |
| SSAHA[68] | https://www.sanger.ac.uk/science/tools/ssaha | 2001 | DNA | hashing | Y | N | N | NW | N | 500 |
| Pattern Hunter[69]–[72] | https://www.bioinfor.com/ | 2002 | DNA | hashing | Y | Y | Y | Non-DP heuristic | N | 500 |
| BLAT[35] | https://genome.ucsc.edu/cgi-bin/hgBlat | 2002 | DNA | hashing | Y | N | Y | Non-DP heuristic | N | 500 |
| BLASTZ[36] | https://www.bx.psu.edu/miller_lab/ | 2003 | DNA | hashing | Y | N | N | SW | Y | 3000 |
| C4[73] | https://github.com/nathanweeks/exonerate | 2005 | DNA | hashing | Y | N | Y | Sparse DP | N | N/A |
| GMAP[74] | https://github.com/juliangehring/GMAP-GSNAP | 2005 | DNA | hashing | N | N | Y | NW | N | N/A |
| BWT-SW[75] | https://github.com/mruffalo/bwt-sw | 2008 | DNA | BWT | Y | N | N | SW | N | 2000 |
| MAQ[41] | http://maq.sourceforge.net/maq-man.shtml | 2008 | DNA | hashing | Y | Y | N | SW | N | 63 |
| RMAP[42] | https://github.com/smithlabcode/rmap | 2008 | DNA | hashing | Y | N | N | HD | N | 36 |
| SOAP[76] | https://github.com/ShujiaHuang/SOAPaligner | 2008 | DNA | hashing | Y | N | N | Non-DP heuristic | N | 50 |
| SOCS[77] | http://socs.biology.gatech.edu/ | 2008 | DNA | hashing | Y | N | N | Rabin-Karp Algorithm | N | 35 |
| SeqMap[44] | http://www-personal.umich.edu/~jianghui/seqmap/ | 2008 | DNA | hashing | Y | N | N | Non-DP Heuristic | N | 30 |
| ZOOM[43] | http://www.bioinfor.com/zoom-1-3-gui-release-next-gen-seq/ | 2008 | DNA | hashing | Y | Y | N | SW | N | 36 |
| QPALMA[78,79] | http://www.raetschlab.org/suppl/qpalma | 2008 | RNA-Seq | suffix array | Y | N | Y | SW | Y | 36 |
| BRAT[80] | http://compbio.cs.ucr.edu/brat/ | 2009 | BS-Seq | hashing | Y | N | N | HD | N | 26 |
| BSMAP[81] | https://github.com/genome-vendor/bsmap | 2009 | BS-Seq | hashing | Y | N | N | HD | N | 32 |
| BFAST[82] | https://github.com/nh13/BFAST/ | 2009 | DNA | hashing | N | Y | N | SW | N | 55 |
| BWA[83] | https://github.com/lh3/bwa | 2009 | DNA | BWT-FM | N | N | N | Semi-Global | N | 125 |
| Bowtie[47] | http://bowtie-bio.sourceforge.net/manual.shtml | 2009 | DNA | BWT-FM | Y | N | N | HD | N | 76 |
| CloudBurst[84] | https://sourceforge.net/projects/cloudburst-bio/ | 2009 | DNA | hashing | Y | N | N | Landau-Vishkin | N | 36 |
| GNUMAP[85] | https://github.com/byucsl/gnumap | 2009 | DNA | hashing | Y | N | Y | NW | N | 36 |
| Genome Mapper[62] | http://1001genomes.org/software/genomemapper_singleref.html | 2009 | DNA | hashing | Y | N | Y | NW | N | 200 |
| MOM[86] | https://github.com/hugheaves/MOM | 2009 | DNA | hashing | Y | N | N | HD | N | 40 |
| PASS[87] | http://pass.cribi.unipd.it/cgi-bin/pass.pl | 2009 | DNA | hashing | Y | N | Y | NW | N | 32 |
| PerM[88] | https://code.google.com/archive/p/perm/downloads | 2009 | DNA | hashing | Y | Y | N | HD | N | 47 |
| RazerS[89] | https://github.com/seqan/seqan/tree/master/apps/razers | 2009 | DNA | hashing | Y | Y | Y | Myers Bit-Vector | N | 76 |
| SHRiMP[90] | http://compbio.cs.toronto.edu/shrimp/ | 2009 | DNA | hashing | N | N | N | SW | N | 35 |
| SOAP2[91] | https://github.com/ShujiaHuang/SOAPaligner | 2009 | DNA | BWT-FM | Y | N | N | SW | N | 44 |
| Slider[92] | http://www.bcgsc.ca/platform/bioinfo/software/slider | 2009 | DNA | hashing | Y | N | N | HD | N | 36 |
| segemehl[93] | https://www.bioinf.uni-leipzig.de/Software/segemehl/ | 2009 | DNA | suffix array | N | N | Y | SW | N | 35 |
| TopHat[94] | https://ccb.jhu.edu/software/tophat/index.shtml | 2009 | RNA-Seq | BWT-FM | Y | N | N | HD | Y | 42 |
| BS-Seeker[95] | http://pellegrini-legacy.mcdb.ucla.edu/bs_seeker/BS_Seeker.html | 2010 | BS-Seq | BWT-FM | Y | N | N | HD | Y | 36 |
| BWA-SW[83] | https://github.com/lh3/bwa | 2010 | DNA | BWT-FM | N | N | N | SW | N | 10000 |
| GASSST[70] | http://www.irisa.fr/symbiose/projects/gassst/ | 2010 | DNA | hashing | Y | Y | Y | Semi-Global | N | 500 |
| GSNAP[72] | https://github.com/juliangehring/GMAP-GSNAP | 2010 | DNA | hashing | Y | N | Y | Non-DP Heuristic | N | 100 |

| SMALT[96] | https://github.com/rcallahan/smalt | 2010 | DNA | hashing | Y | N | Y | SW | N | 150 |
|---|---|---|---|---|---|---|---|---|---|---|
| Slider II[97] | http://www.bcgsc.ca/platform/bioinfo/software/SliderII | 2010 | DNA | hashing | Y | N | N | HD | Y | 42 |
| VMATCH[98] | http://www.vmatch.de/ | 2010 | DNA | suffix array | Y | N | Y | SW | Y | N/A |
| mrsFAST[99] | https://github.com/sfu-compbio/mrsfast | 2010 | DNA | hashing | Y | N | N | HD | N | 100 |
| MapSplice[100] | https://github.com/LiuBioinfo/MapSplice | 2010 | RNA-Seq | BWT-FM | Y | N | N | HD | Y | 100 |
| MicroRazerS[101] | https://github.com/seqan/seqan/tree/master/apps/micro_razers | 2010 | RNA-Seq | hashing | Y | N | Y | HD | N | 36 |
| SpliceMap[102] | http://web.stanford.edu/group/wonglab/SpliceMap/ | 2010 | RNA-Seq | hashing | Y | N | N | HD | Y | 50 |
| Supersplat[61] | http://mocklerlab.org/tools/1/manual | 2010 | RNA-Seq | hashing | N | N | N | NA | N | 36 |
| Bismark[103] | https://github.com/FelixKrueger/Bismark | 2011 | BS-Seq | BWT-FM | Y | N | Y | SW & NW | Y | 50 |
| LAST[51] | http://last.cbrc.jp/ | 2011 | DNA/BS-Seq/RNA | suffix array | N | Y | N | SW & NW | N | 105 |
| DynMap[104] | https://dl.acm.org/citation.cfm?id=2147845&dl=ACM&coll=DL | 2011 | DNA | hashing | Y | N | N | NW | N | 52 |
| SHRiMP2[40] | http://compbio.cs.toronto.edu/shrimp/ | 2011 | DNA | hashing | Y | Y | Y | SW | N | 75 |
| SNAP[105] | http://snap.cs.berkeley.edu/ | 2011 | DNA | hashing | Y | N | N | NW | N | 10000 |

| Stampy[106] | https://www.well.ox.ac.uk/project-stampy | 2011 | DNA | hashing | Y | N | N | NW | N | 4500 〖F〗〖SEP〗 |
|---|---|---|---|---|---|---|---|---|---|---|
| TMAP | https://github.com/iontorrent/TS/tree/master/Analysis/TMAP | 2011 | DNA | BWT-FM | N | N | Y | SW | N | N/A |
| X-Mate[107] | http://grimmond.imb.uq.edu.au/X-MATE/ | 2011 | DNA | hashing | N | N | N | Non-DP Heuristic | N | 50 |
| SOAPSplice[108] | http://soap.genomics.org.cn/soapsplice.html | 2011 | RNA-Seq | BWT-FM | Y | N | N | Non-DP Heuristic | N | 150 |
| BRAT-BW[80] | http://compbio.cs.ucr.edu/brat/ | 2012 | BS-Seq | BWT-FM | N | N | N | HD | N | 62 |
| BLASR[109] | https://github.com/mchaisso/blasr/ | 2012 | DNA | suffix array | N | N | Y | NW | N | 8000 |
| Batmis[110] | https://code.google.com/archive/p/batmis/ | 2012 | DNA | BWT-ST | Y | N | N | HD | N | 100 |
| Bowtie2[111] | http://bowtie-bio.sourceforge.net/bowtie2 | 2012 | DNA | BWT-FM | Y | N | Y | SW & NW | N | 400 |
| GEM[112] | https://github.com/smarco/gem3-mapper | 2012 | DNA | BWT-FM | N | N | Y | SW & NW | N | 150 |
| RazerS3[113] | https://github.com/seqan/seqan/tree/master/apps/razers3 | 2012 | DNA | hashing | Y | Y | Y | Banded Myers Bit Vector | N | 800 |
| SeqAlto[114] | https://web.stanford.edu/group/wonglab/seqalto/ | 2012 | DNA | hashing | Y | N | N | NW | N | 200 |
| SplazerS[115] | https://github.com/seqan/seqan/blob/master/apps/splazers/README | 2012 | DNA | hashing | Y | N | Y | Banded Myers Bit Vector | N | 150 |
| WHAM[116] | http://pages.cs.wisc.edu/~jignesh/wham/ | 2012 | DNA | hashing | Y | N | N | NW | N | 74 |
| YAHA[117] | https://github.com/GregoryFaust/yaha | 2012 | DNA | hashing | Y | N | Y | SW | N | 10000 |
| OSA[118] | http://www.arrayserver.com/wiki/index.php?title=OSA | 2012 | RNA-Seq | hashing | Y | N | N | NA | N | 100 |
| Passion[119] | https://trac.nbic.nl/passion/ | 2012 | RNA-Seq | hashing | Y | N | Y | SW | Y | 75 |
| BS-Seeker2[120] | https://github.com/BSSeeker/BSseeker2 | 2013 | BS-Seq | BWT-FM | Y | N | Y | SW & NW | Y | 250 |
| Subread[121] | http://subread.sourceforge.net/ | 2013 | DNA/RNA-Seq | hashing | Y | Y | Y | SW | N | 202 |
| BWA-MEM[122] | https://github.com/lh3/bwa | 2013 | DNA | BWT-FM | N | N | Y | SW & NW | N | 650 |
| Masai[57] | http://www.seqan.de/projects/masai | 2013 | DNA | suffix tree | N | N | N | Banded Myers Bit Vector | N | 150 |
| NextGenMap[123] | http://cibiv.github.io/NextGenMap/ | 2013 | DNA | hashing | Y | N | N | SW & NW | N | 250 |
| SRmapper[124] | http://www.umsl.edu/~wongch/software.html | 2013 | DNA | hashing | Y | N | N | HD | N | 100 |
| mrFAST[125] | https://github.com/BilkentCompGen/mrfast | 2013 | DNA | hashing | Y | N | N | Semi-Global | N | 180 |
| CRAC[126] | http://crac.gforge.inria.fr/ | 2013 | RNA-Seq | BWT-FM | Y | N | N | Non-DP Heuristic | N | 200 |
| STAR[127] | https://github.com/alexdobin/STAR | 2013 | RNA-Seq | suffix array | N | N | N | SW | N | 5000 |
| TopHat2[128] | https://ccb.jhu.edu/software/tophat/index.shtml | 2013 | RNA-Seq | BWT-FM | Y | N | Y | SW & NW | Y | 101 |
| Subjunc[89] | http://subread.sourceforge.net/ | 2013 | RNA-seq | hashing | Y | Y | Y | NW | N | 202 |
| BWA-PSSM[129] | http://bwa-pssm.binf.ku.dk/ | 2014 | DNA | BWT-FM | Y | N | N | SW | Y | 100 |
| CUSHAW3[130] | http://cushaw3.sourceforge.net/homepage.htm#latest | 2014 | DNA | BWT-FM | Y | N | Y | SW & Semi-Global | N | 100 |

| Hobbes 2[131] | https://hobbes.ics.uci.edu/download.shtml | 2014 | DNA | hashing | Y | N | Y | Banded Myers Bit Vector | N | 100 |
|---|---|---|---|---|---|---|---|---|---|---|
| MOSAIK[132] | https://github.com/wanpinglee/MOSAIK | 2014 | DNA | hashing | Y | N | N | SW | N | 100 |
| hpg-Aligner[133] | https://github.com/opencb/hpg-aligner | 2014 | DNA | suffix array | N | N | Y | SW | N | 5000 |
| mrsFAST-Ultra[134] | https://github.com/sfu-compbio/mrsfast | 2014 | DNA | hashing | Y | N | N | HD | N | 100 |
| JAGuaR[135] | http://www.bcgsc.ca/platform/bioinfo/software/jaguar | 2014 | RNA-Seq | BWT-FM | Y | N | N | SW | Y | 100 |
| Context Map 2[136] | http://www.bio.ifi.lmu.de/ContextMap | 2015 | RNA-Seq | BWT-FM | Y | N | Y | SW & NW | Y | 76 |
| HISAT[63] | http://www.ccb.jhu.edu/software/hisat/index.shtml | 2015 | RNA-Seq | BWT-FM | Y | N | N | Non-DP Heuristic | N | 100 |
| ERNE 2[52] | http://erne.sourceforge.net/ | 2016 | DNA/BS-Seq | BWT-FM + hashing | Y | N | N | HD | N | 100 |
| GraphMap[137] | https://github.com/isovic/graphmap | 2016 | DNA | hashing | Y | Y | Y | Semi-global | N | 9000 |
| NanoBLASTer[138] | https://github.com/ruhulsbu/NanoBLASTer | 2016 | DNA | hashing | Y | N | Y | NW | N | 7040 |
| minimap[139] | https://github.com/lh3/minimap | 2016 | DNA | hashing | Y | N | N | N/A | N | 13000 |
| rHAT[140] | https://github.com/dfguan/rHAT | 2016 | DNA | hashing | Y | N | Y | SW | N | 8000 |
| KART[141] | https://github.com/hsinnan75/KART | 2017 | DNA | BWT-FM | N | N | N | NW | N | 7118 |
| LAMSA[53] | https://github.com/hitbc/LAMSA | 2017 | DNA | BWT-FM + hashing | Y | N | Y | Sparse DP | Y | 100000 |
| DART[142] | https://github.com/hsinnan75/DART | 2017 | RNA-Seq | BWT-FM | N | N | Y | NW | N | 251 |
| minimap2[143] | https://github.com/lh3/minimap2 | 2018 | DNA/RNA-Seq | hashing | Y | N | Y | NW | N | 11628 |
| DREAM-Yara[64] | https://gitlab.com/pirovc/dream_yara/ | 2018 | DNA | BWT-FM | Y | N | N | Banded Myers Bit Vector | Y | 150 |
| MUMmer4[144] | https://github.com/mummer4/mummer | 2018 | DNA | suffix array | Y | N | Y | SW | Y | 7821 |
| NGMLR[145] | https://github.com/philres/ngmlr | 2018 | DNA | hashing | Y | N | Y | SW | N | 50000 |
| lordFAST[54] | https://github.com/vpc-ccg/lordfast | 2018 | DNA | BWT-FM + hashing | N | N | Y | SW & NW | N | 35489 |
| BatMeth2[146] | https://github.com/GuoliangLi-HZAU/BatMeth2/ | 2019 | BS-Seq | BWT-FM | Y | N | Y | SW & NW | N | 125 |
| GraphMap2[147] | https://github.com/lbcb-sci/graphmap2 | 2019 | DNA/RNA-Seq | hashing | Y | Y | Y | Semi-global | N | 9000 |
| Magic-BLAST[148] | https://github.com/ncbi/magicblast | 2019 | DNA/RNA-Seq | hashing | Y | N | N | Non-DP Heuristic | N | 90000 |
| BWA-MEM2[149] | https://github.com/bwa-mem2/bwa-mem2 | 2019 | DNA | BWT-FM | N | N | Y | SW | N | 650 |
| HISAT2[150] | https://ccb.jhu.edu/software/hisat2/index.shtml | 2019 | DNA | BWT-FM | Y | N | N | Non-DP Heuristic | N | 100 |
| deSALT[151] | https://github.com/hitbc/deSALT | 2019 | RNA-seq | hashing | Y | N | N | SW | N | 8000 |
| conLSH[152] | https://www.dropbox.com/s/3jcu4i240kyu2tc/source%20code%20conLSH_bi | 2020 | DNA | hashing | Y | N | Y | Sparse DP | N | 8000 |

| | o.tar.gz?dl=0 | | | | | | | | | | |
|---|---|---|---|---|---|---|---|---|---|---|---|

**Table 2. Advantages and limitations of read alignment algorithms.** We compare the ease of implementing each algorithm ('Easy to implement'). We also record whether the algorithm allows for an exact and/or inexact match ('Search for exact/inexact match'). The use of spaced seeds enables searching for inexact match using a hash table. We also compare the size of genome index (indicated in column "Index size"), the speed of seed query (indicated in column "Seed query speed"), and the possibility to vary the length of the seed ("Seed length").

|  | Hashing | Suffix tree and BWT-FM |
|---|---|---|
| Easy to implement | Yes | No |
| Search for exact/inexact match | Exact | Exact and Inexact |
| Index size | Large | Compressed (small) |
| Indexing time | Small | Large |
| Seed query speed | O(1), fast | Slow |
| Seed length | Fixed length per index | Can be fixed or variable |

**a. Seed extraction from reference genome**

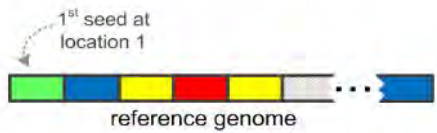

**b. Seed indexing using suffix tree or hash table**

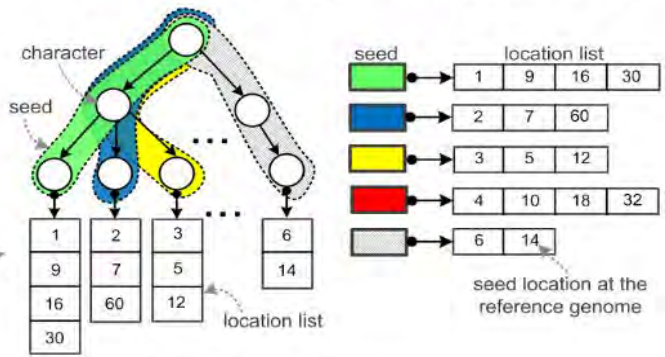

**c. Seed extraction from reads**

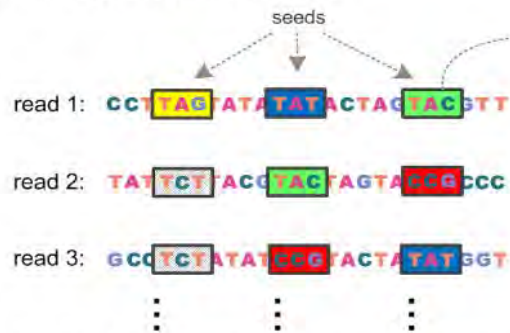

**d. Seed querying and filtering**

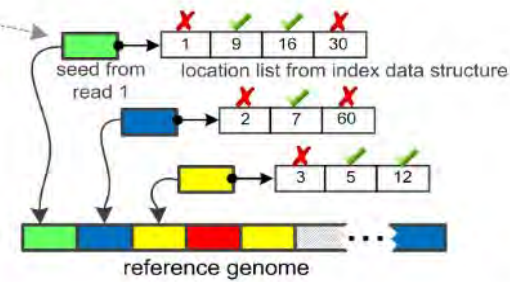

**e. Seed chaining and pre-alignment filtering**

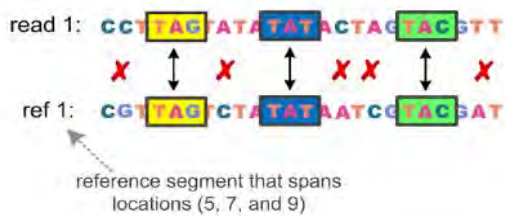

**f. Alignment verification**

|   | C | G | T | T | A | G | T | C | T | A | ... |
|---|---|---|---|---|---|---|---|---|---|---|---|
|   | 0 | 0 | 0 | 0 | 0 | 0 | 0 | 0 | 0 | 0 | 0 |
| C | 0 | 2 | 2 | 2 | 2 | 2 | 2 | 2 | 2 | 2 | 2 |
| C | 0 | 2 | 3 | 3 | 3 | 3 | 3 | 4 | 4 | 4 | 4 |
| T | 0 | 2 | 3 | 5 | 5 | 5 | 5 | 5 | 6 | 6 | 6 |
| T | 0 | 2 | 3 | 5 | 7 | 7 | 7 | 7 | 7 | 7 | 7 |
| A | 0 | 3 | 3 | 5 | 7 | 9 | 9 | 9 | 9 | 9 | 9 |
| G | 0 | 2 | 4 | 5 | 7 | 9 | 11 | 11 | 11 | 11 | 11 |
| T | 0 | 2 | 4 | 6 | 7 | 9 | 11 | 13 | 13 | 13 | 13 |
| A | 0 | 2 | 3 | 5 | 7 | 9 | 11 | 13 | 14 | 14 | 15 |
| T | 0 | 2 | 4 | 6 | 8 | 9 | 11 | 13 | 14 | 16 | 16 |

.bam/.sam file contains necessary alignment information (e.g., type, location, and number of each edit)

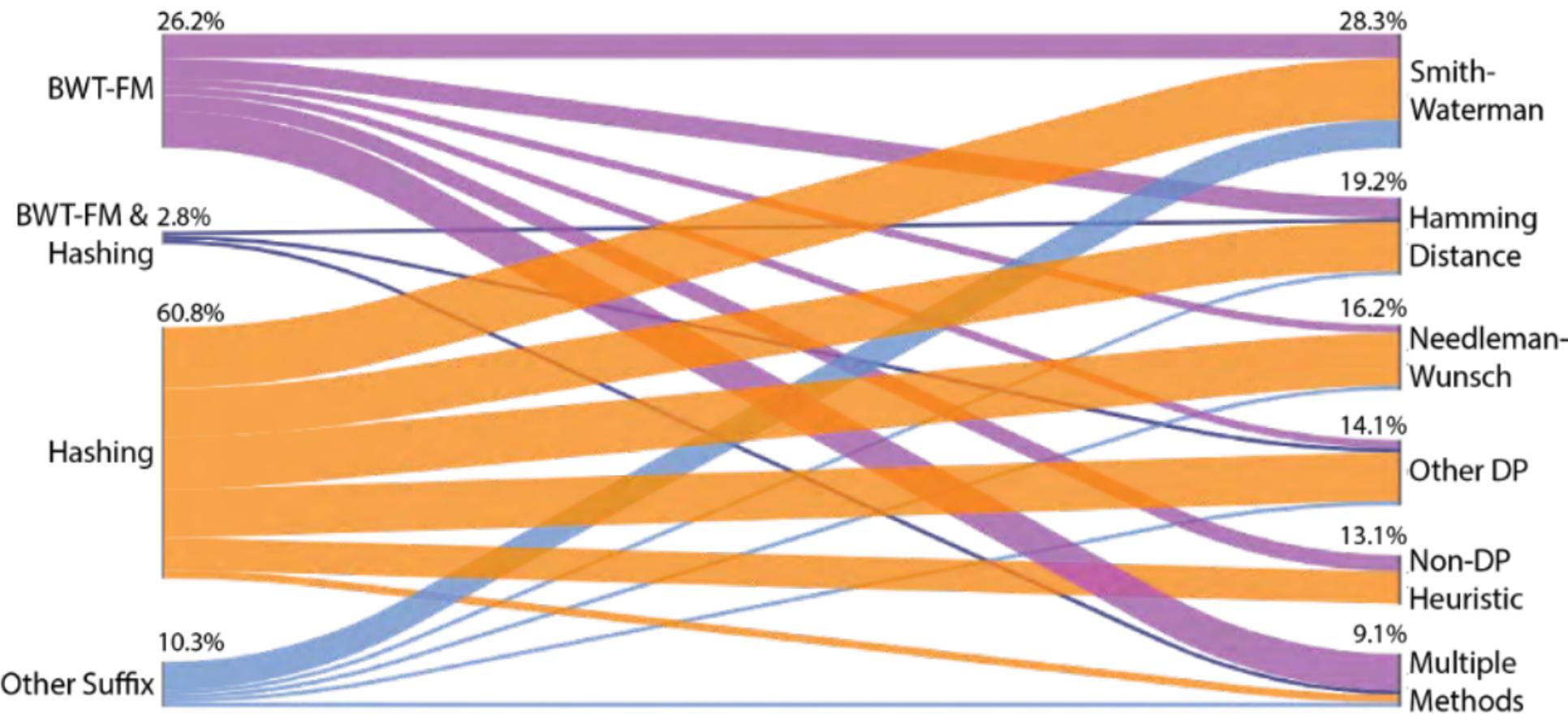

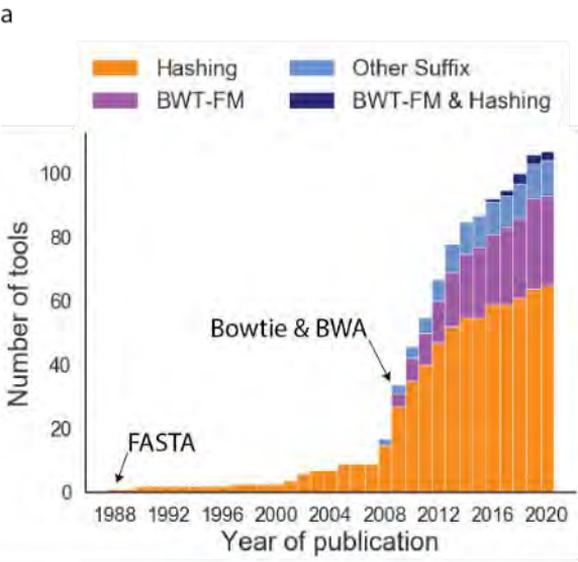

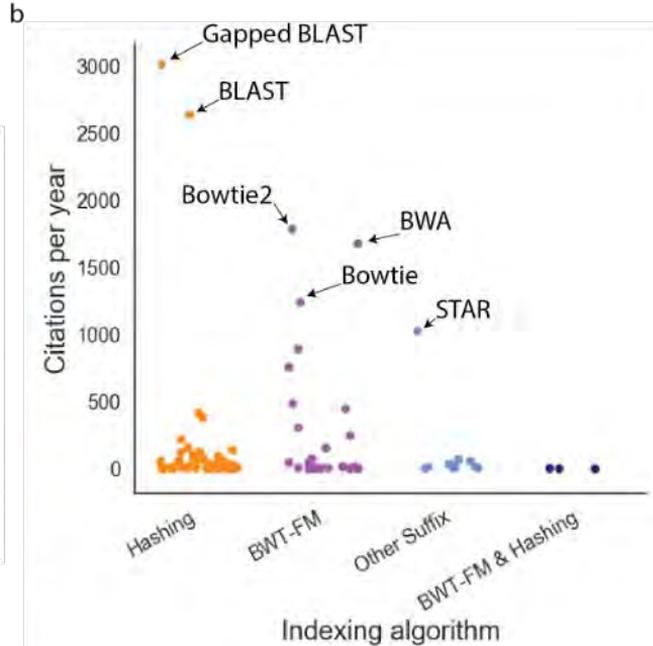

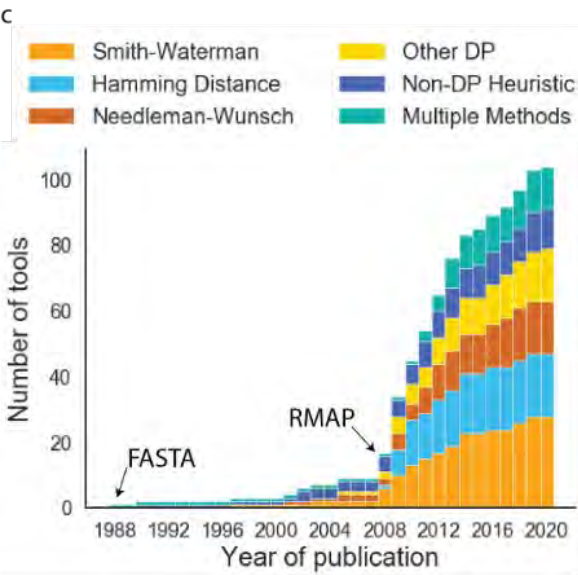

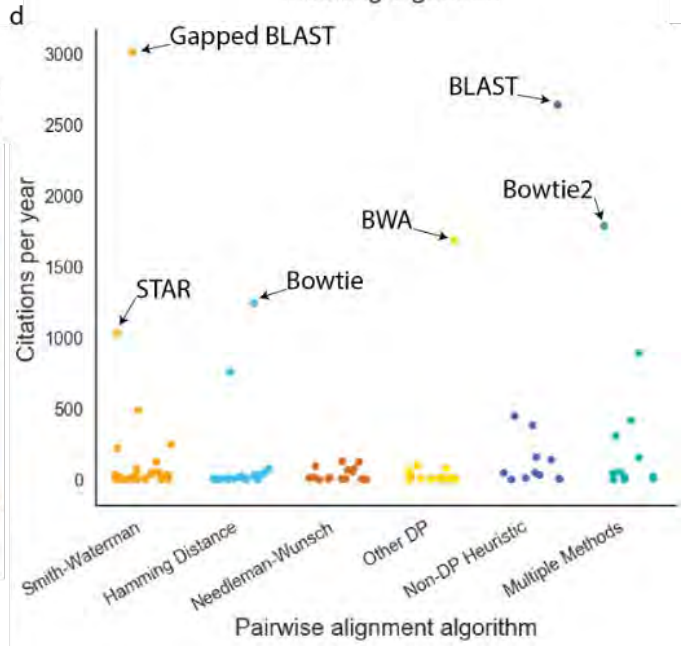

**Technology dictates algorithms: Recent developments in read alignment**

Mohammed Alser[1,2,†], Jeremy Rotman[3,†], Kodi Taraszka[3], Huwenbo Shi[4,5], Pelin Icer Baykal[6], Harry Taegyun Yang[3,7], Victor Xue[3], Sergey Knyazev[6], Benjamin D. Singer[8,9,10], Brunilda Balliu[11], David Koslicki[12,13,14], Pavel Skums[6], Alex Zelikovsky[6,15], Can Alkan[2,16], Onur Mutlu[1,#], Serghei Mangul[17,#,*]

**Supplementary Table 1. Genome index size across three read alignment tools.**

| Tool | Version | Index Size | Indexing Time |
|------|---------|------------|---------------|
| mrFAST | 2.2.5 | 16.5 GB | 1202 seconds |
| minimap2 | 0.12.7 | 7.2 GB | 200 seconds |
| BWA-MEM | 0.7.17 | 4.7 GB | 2998 seconds |

**Supplementary Table 2. Tools available in the bioconda package manager. Here we have included only the tools that are primarily built for DNA read alignment.**

| Software tool | Version | Publication | Conda command |
|---|---|---|---|
| Bowtie2 | 2.2.5 | [111] | conda install -c bioconda bowtie2 |
| Bowtie | 0.12.7 | [47] | conda install -c bioconda bowtie |
| BWA | 0.7.17 | [83] | conda install -c bioconda bwa |
| GSNAP | 2018-03-25 | [72] | conda install -c compbiocore gsnap |
| HISAT2 | 2.1.0 | [254] | conda install -c bioconda hisat2 |
| LAST | 963 | [51] | conda install -c bioconda last |
| minimap2 | 2.12-r827 | [143] | conda install -c bioconda minimap2 |
| RMAP | 2.1 | [128] | conda install -c bioconda rmap |
| SMALT | 0.7.6 | [96] | conda install -c bioconda smalt |
| SNAP | 1.0beta.23 | [105] | conda install -c bioconda snap-aligner |
| Subread | v1.6.2 | [121] | conda install -c bioconda subread |

**Supplementary Table 3. Fixed effect size estimates, standard errors (se), test statistics, and p-values for the effect of the listed variables on the expected CPU run time.** The parameters were estimated using the Gamma generalized linear mixed model in Equation (1). "Variable name: Level 1 vs Level 2" indicates that Level 1 is the reference level and the coefficient quantifies the increase / decrease in expected CPU run time for Level 2 over Level 1.

| | Estimate | se | t stat | p-value |
|---|---|---|---|---|
| Intercept | 0.19 | 0.23 | 0.81 | 4.2e-01 |
| Year of publication | -0.7 | 0.09 | -7.96 | 1.7e-15 |
| Chain of seeds: No vs Yes | 1.45 | 0.19 | 7.46 | 8.8e-14 |
| Pairwise alignment: NW vs HD | 1.37 | 0.28 | 4.91 | 9.3e-07 |
| Pairwise alignment: NW vs Non-DP Heuristic | 1.22 | 0.19 | 6.37 | 1.8e-10 |
| Pairwise alignment: NW vs SW | 0.78 | 0.2 | 3.83 | 1.3e-04 |
| Indexing: hashing vs BWT-FM | -0.11 | 0.16 | -0.68 | 5.0e-01 |

**Supplementary Table 4. Likelihood ratio test p-values for the effect of the listed variables on the expected CPU run time.** The parameters (Supplementary Table 3) were estimated using the Gamma generalized linear mixed model in Equation (1). The null Gamma generalized linear mixed model is generated as in Equation (1), but without the variable of interest. Additionally, we ran one LRT between BWT-FM tools and LAST, an aligner that does not use BWT and the FM-index by default.

| Variable of Interest | LRT p-value |
|---|---|
| Indexing | 5.0e-01 |
| Year of publication | 3.7e-11 |
| Pairwise alignment | 3.7e-08 |
| BWT-FM vs LAST | 1.5e-15 |

**Supplementary Table 5. Fixed effect size estimates, standard errors (se), test statistics, and p-values for the effect of the listed variables on the expected RAM usage.** The parameters were estimated using the Gamma generalized linear mixed model in Equation (2). "Variable name: Level 1 vs Level 2" indicates that Level 1 is the reference level and the coefficient quantifies the increase / decrease in expected RAM usage for Level 2 over Level 1.

|  | Estimate | se | t stat | p-value |
|---|---|---|---|---|
| Intercept | 3.41 | 0.51 | 6.67 | 2.2e-02 |
| Year of publication | -0.21 | 0.24 | -0.85 | 4.8e-01 |
| Chain of seeds: No vs Yes | -0.5 | 0.51 | -0.99 | 4.3e-01 |
| Pairwise alignment: NW vs HD | -1.12 | 0.73 | -1.53 | 2.7e-01 |
| Pairwise alignment: NW vs Non-DP Heuristic | 0.2 | 0.5 | 0.4 | 7.3e-01 |
| Pairwise alignment: NW vs SW | -1.11 | 0.54 | -2.06 | 1.8e-01 |
| Indexing: hashing vs BWT-FM | -1.51 | 0.44 | -3.43 | 7.6e-02 |

**Supplementary Table 6. Likelihood ratio test p-values for the effect of the listed variables on the expected RAM usage.** The parameters (Supplementary Table 5) were estimated using the Gamma generalized linear mixed model in Equation (2). The null Gamma generalized linear mixed model is generated as in Equation (2), but without the variable of interest. Additionally, we ran one LRT between BWT-FM tools and LAST, an aligner that does not use BWT and the FM-index by default.

| Variable of Interest | LRT p-value |
|---|---|
| Indexing | 2.2e-03 |
| Year of publication | 4.1e-01 |
| Pairwise alignment | 3.9e-02 |
| BWT-FM vs LAST | 3.2e-77 |

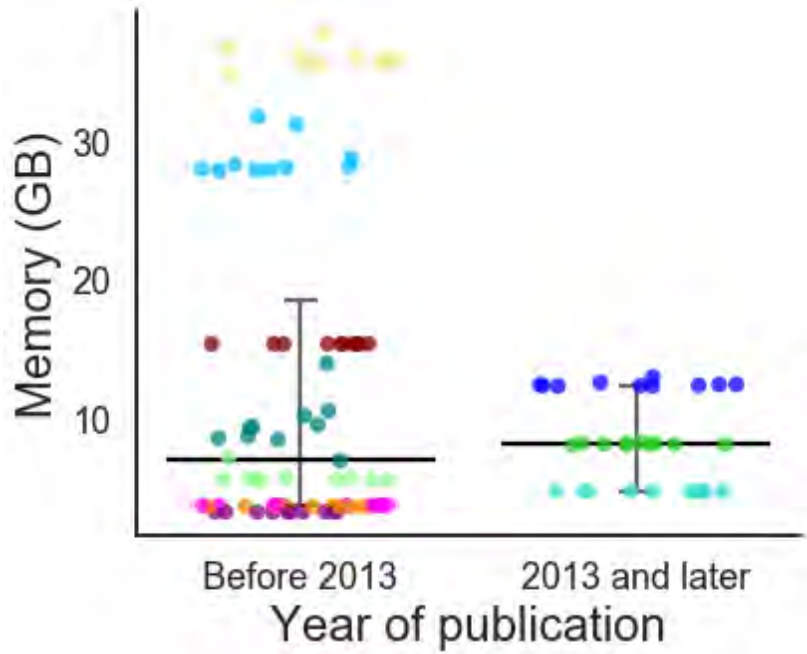

**Supplementary Figure 1. The effect of year of publication on computational resources.** The relative performance (RAM) of the benchmarked aligners grouped by whether the tool was released before or after long read technology was introduced (2013) and colored by individual aligners.

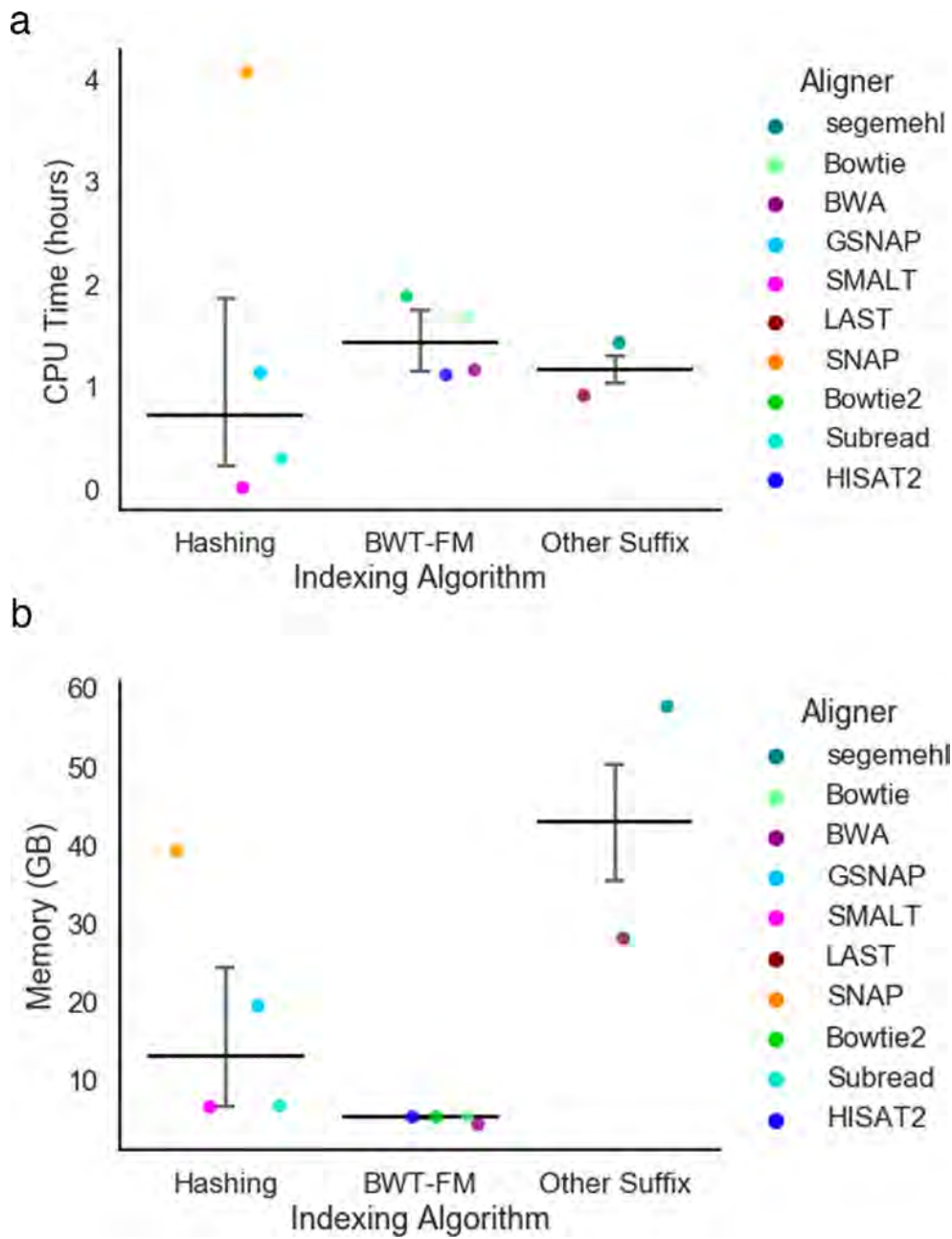

**Supplementary Figure 2. Relative performance of human genome indexing performed by various read alignment tools.** Tools are grouped based on the type of algorithm used for genome indexing. Tools are ordered from oldest (segemehl, 2009) to newest (HISAT2, 2019). (a) CPU time (b) RAM.

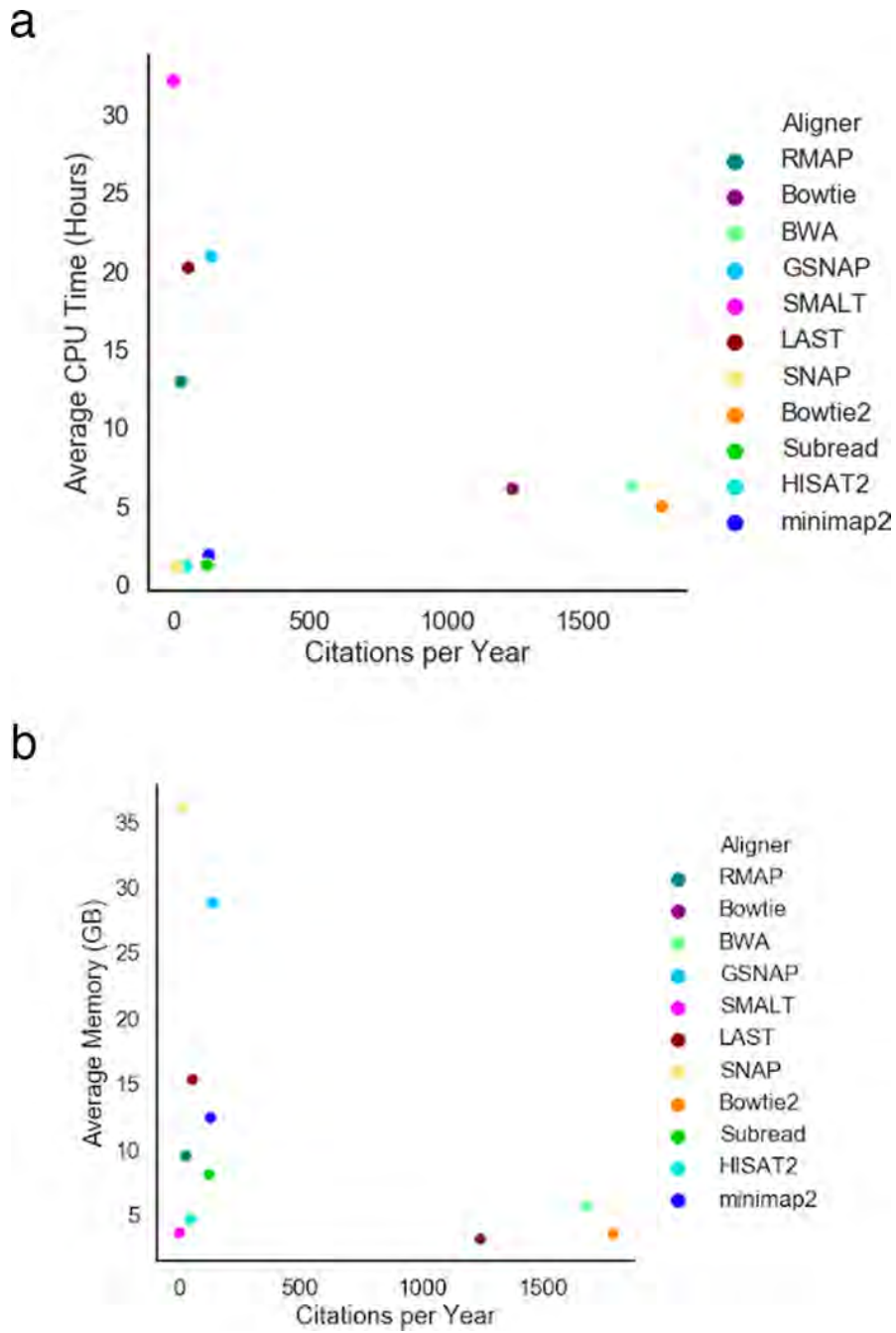

**Supplementary Figure 3. Average relative performance of various read alignment tools plotted against the number of citations the tool's corresponding paper has received yearly since being published.** Tools are ordered from oldest (RMAP, 2008) to newest (minimap2, 2019). (a) CPU time. (b) RAM.

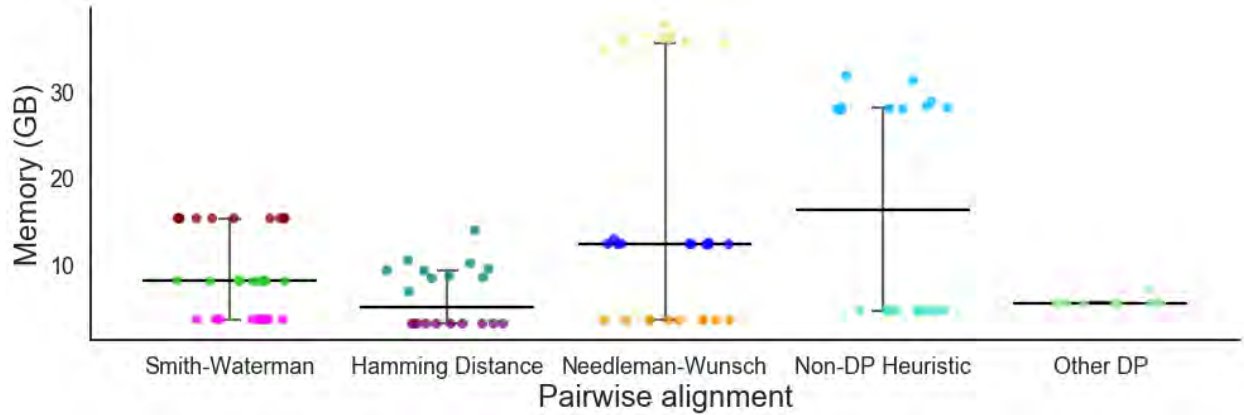

**Supplementary Figure 4. The effect of pairwise alignment algorithms on computational resources.** The relative performance (RAM) of the benchmarked aligners grouped by the algorithm used for pairwise alignment and colored by individual aligners.

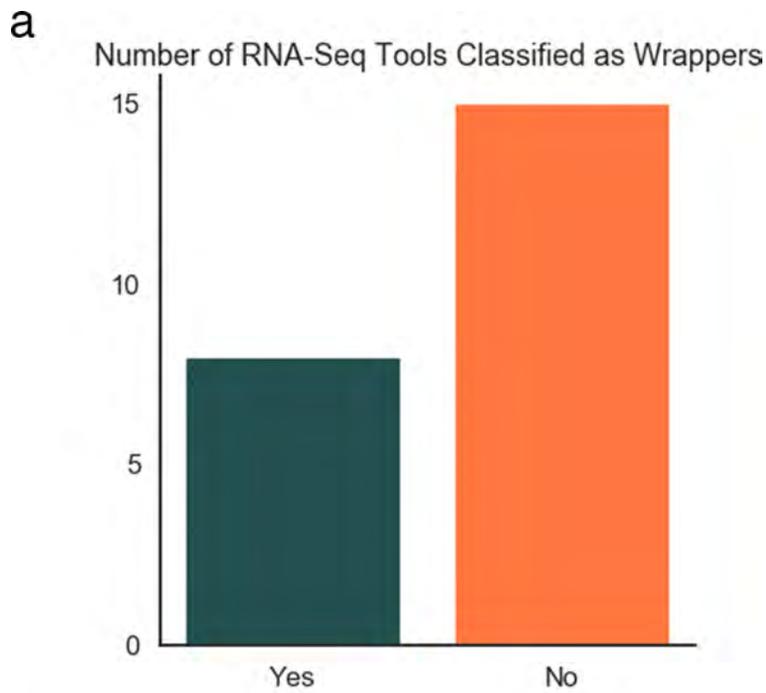

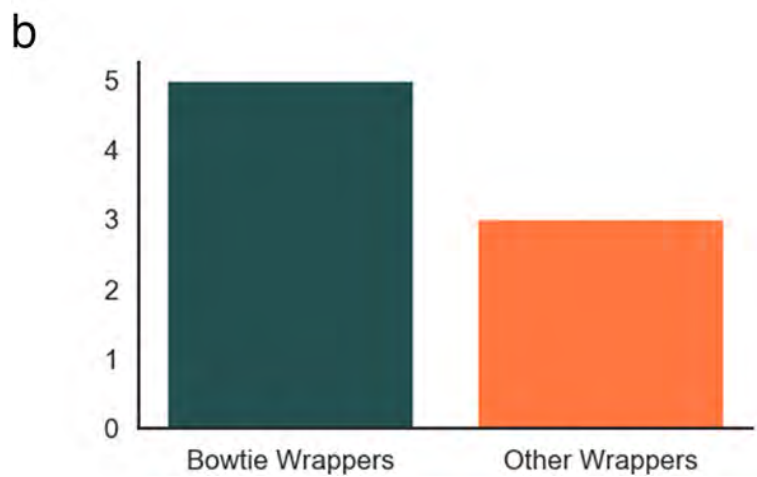

**Supplementary Figure 5.** (a) Bar chart showing the number of surveyed RNA-Seq tools which are wrappers of existing DNA-Seq aligners tools. (b) Bar chart showing the number of surveyed RNA-Seq tools which are wrappers of Bowtie or Bowtie2.

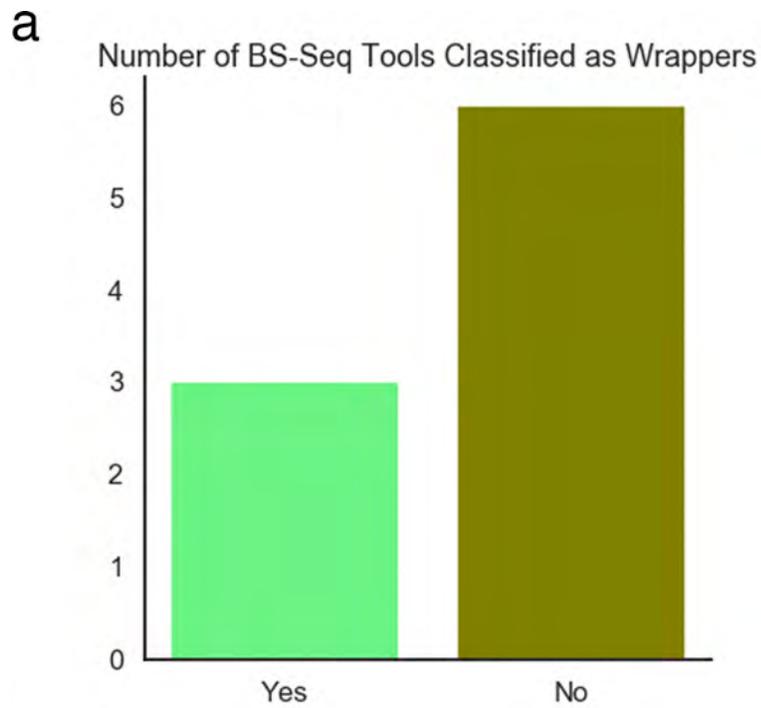

a

Number of BS-Seq Tools Classified as Wrappers

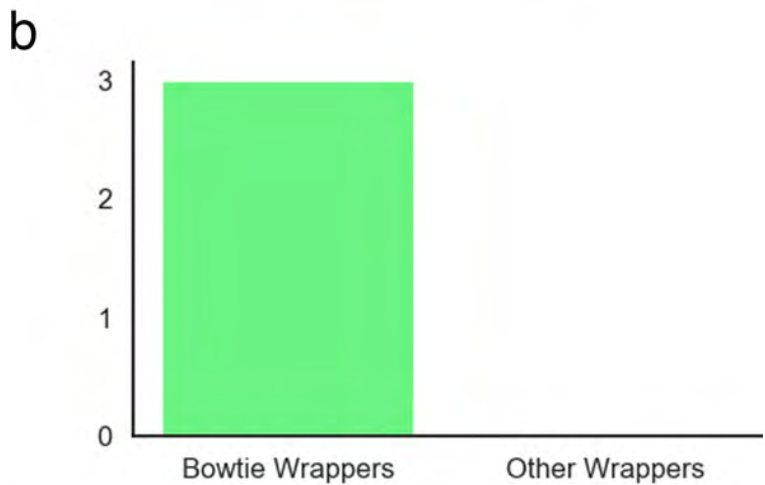

b

**Supplementary Figure 6.** (a) Bar chart showing the number of surveyed BS-Seq tools which are wrappers of existing DNA-Seq aligners tools. (b) Bar chart showing the number of surveyed BS-Seq tools which are wrappers of Bowtie or Bowtie2.

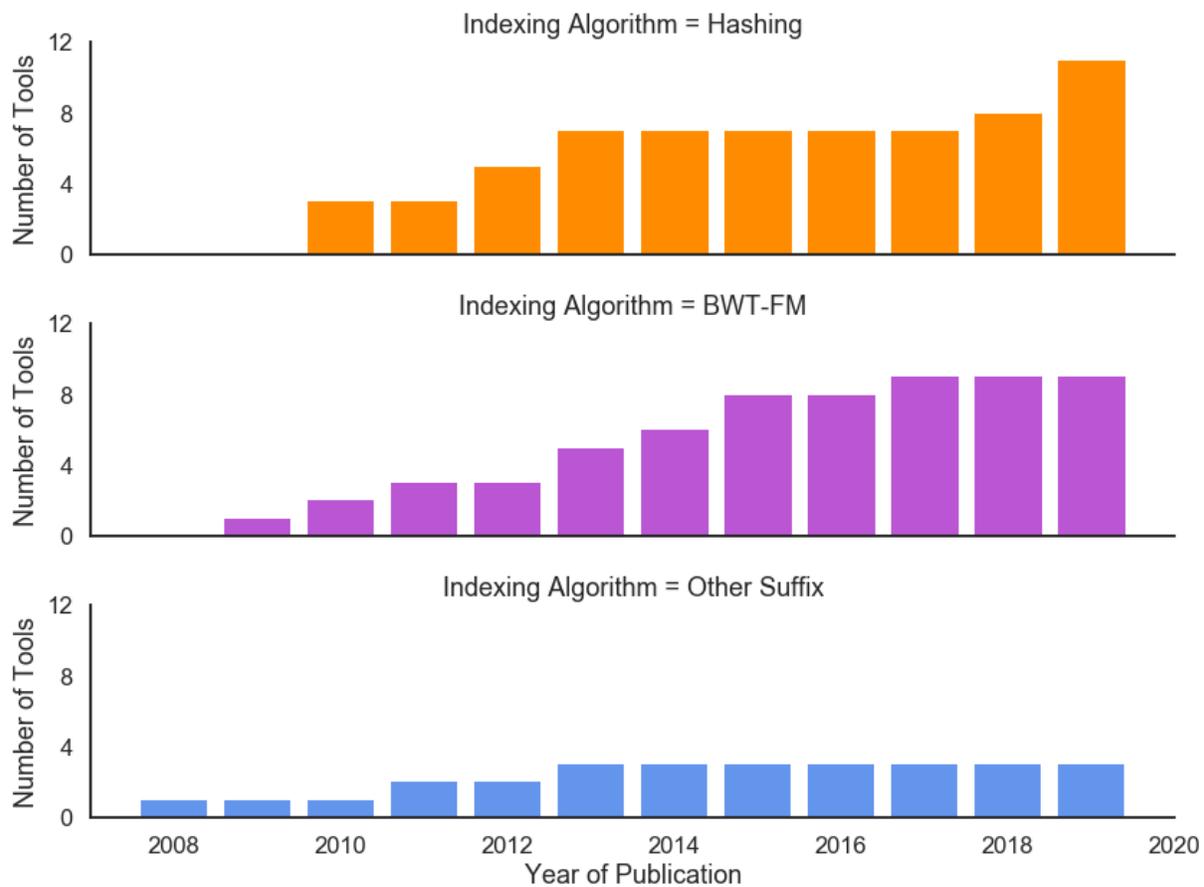

**Supplementary Figure 7. Histogram showing the cumulation of surveyed RNA-Seq tools over time separated by the algorithm used for genome indexing.** This includes both stand alone RNA-Seq tools and wrappers of existing DNA-Seq alignment tools.

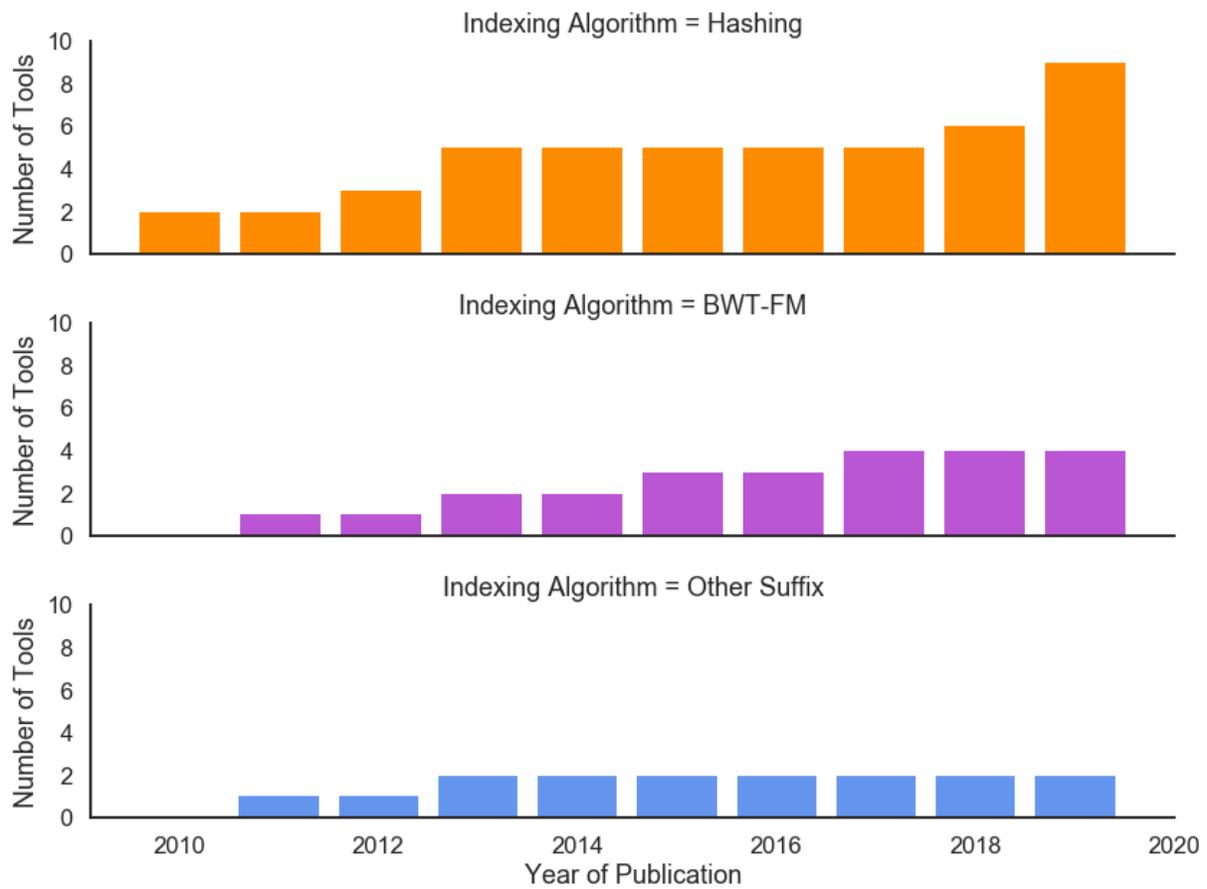

**Supplementary Figure 8. Histogram showing the cumulation of surveyed RNA-Seq tools over time separated by the algorithm used for genome indexing.** Only stand alone RNA-Seq aligners tools are included (not the wrappers of existing DNA-Seq aligners).

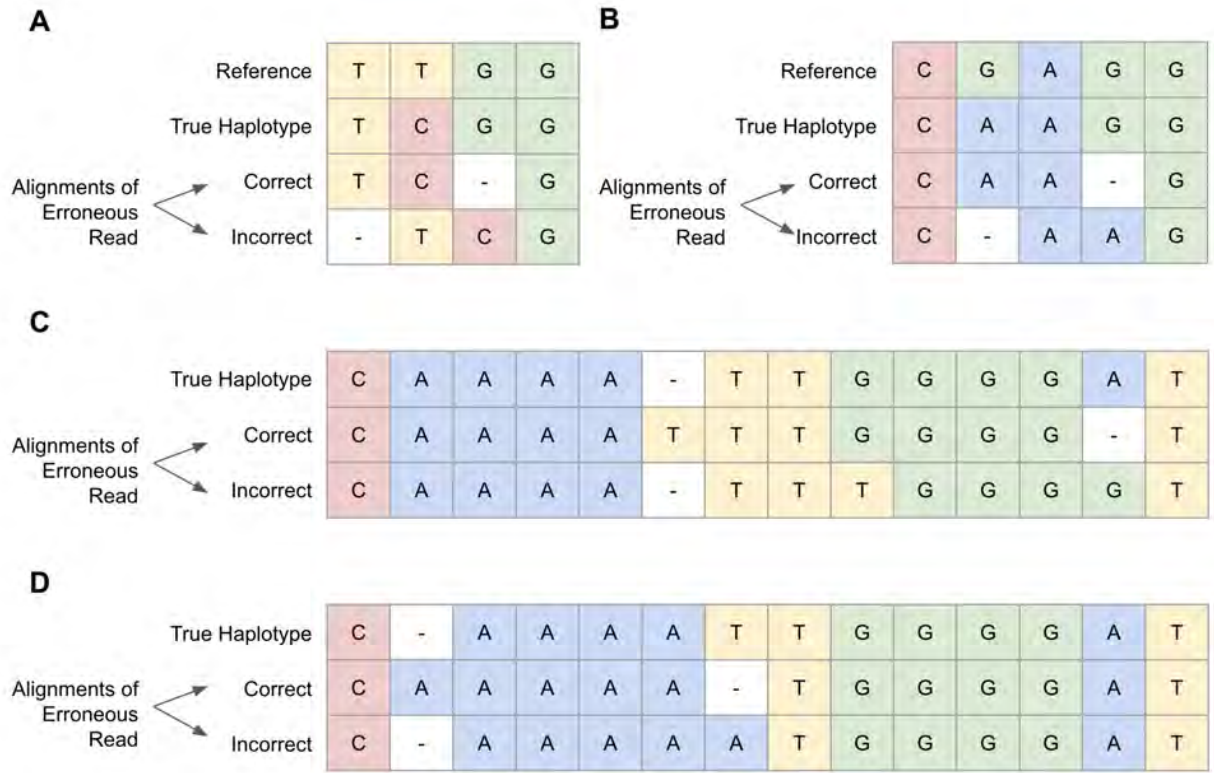

**Supplementary Figure 9. Examples of erroneous alignments of Influenza A virus PacBio sequencing dataset[222].** (A) and (B) Reads come from a true haplotype with the deletion with respect to the reference. Using BWA scoring method, the reads have two different alignments with the optimal score, but only the first alignment is correct. (C) and (D) Correct read alignment with the homopolymer errors should introduce an insertion and a deletion instead of "optimal" two mismatches.

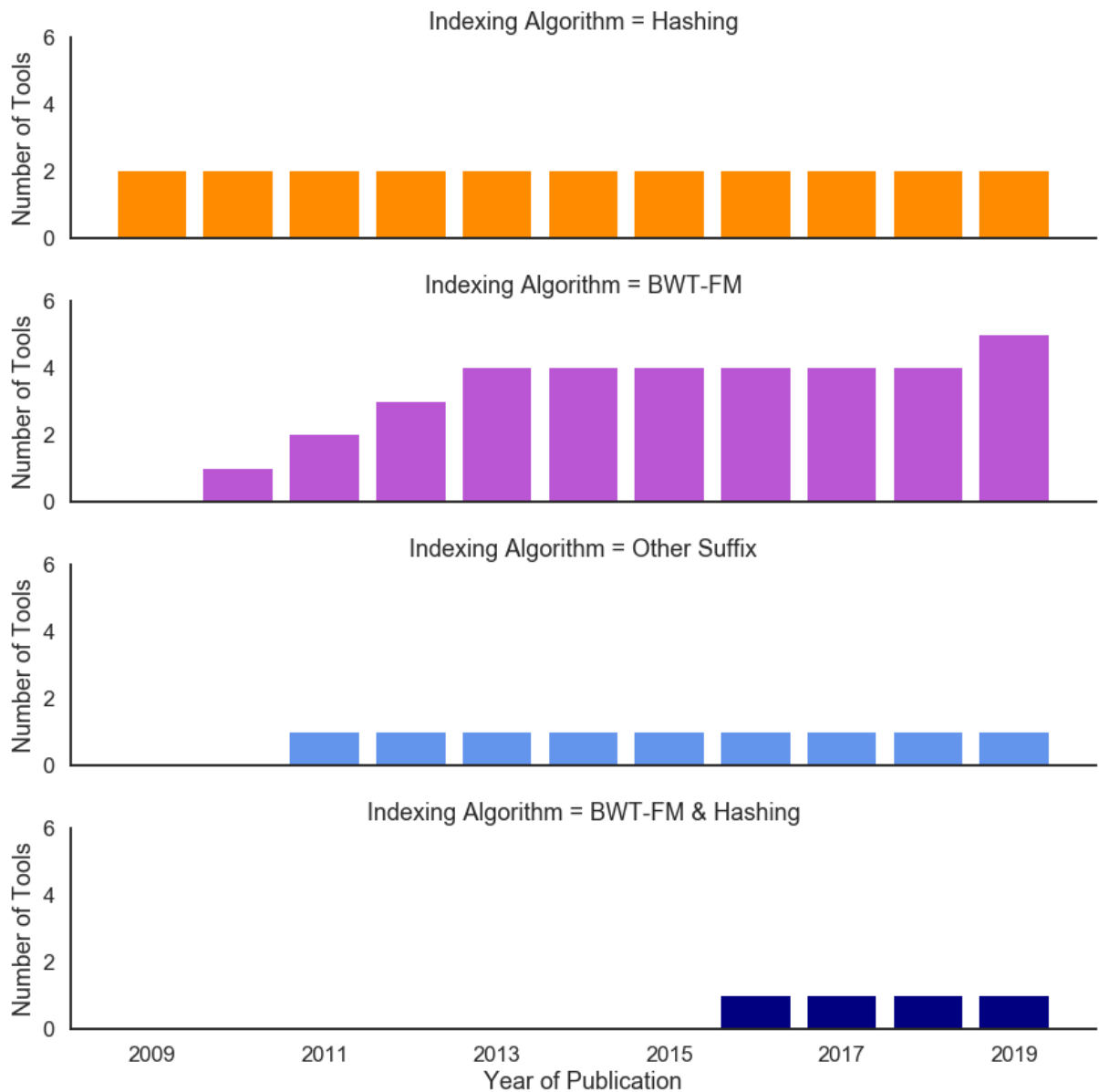

**Supplementary Figure 10. Histogram showing the cumulation of surveyed BS-Seq tools over time separated by the algorithm used for genome indexing.** This includes both stand-alone BS-Seq tools and wrappers of existing DNA-Seq alignment tools.

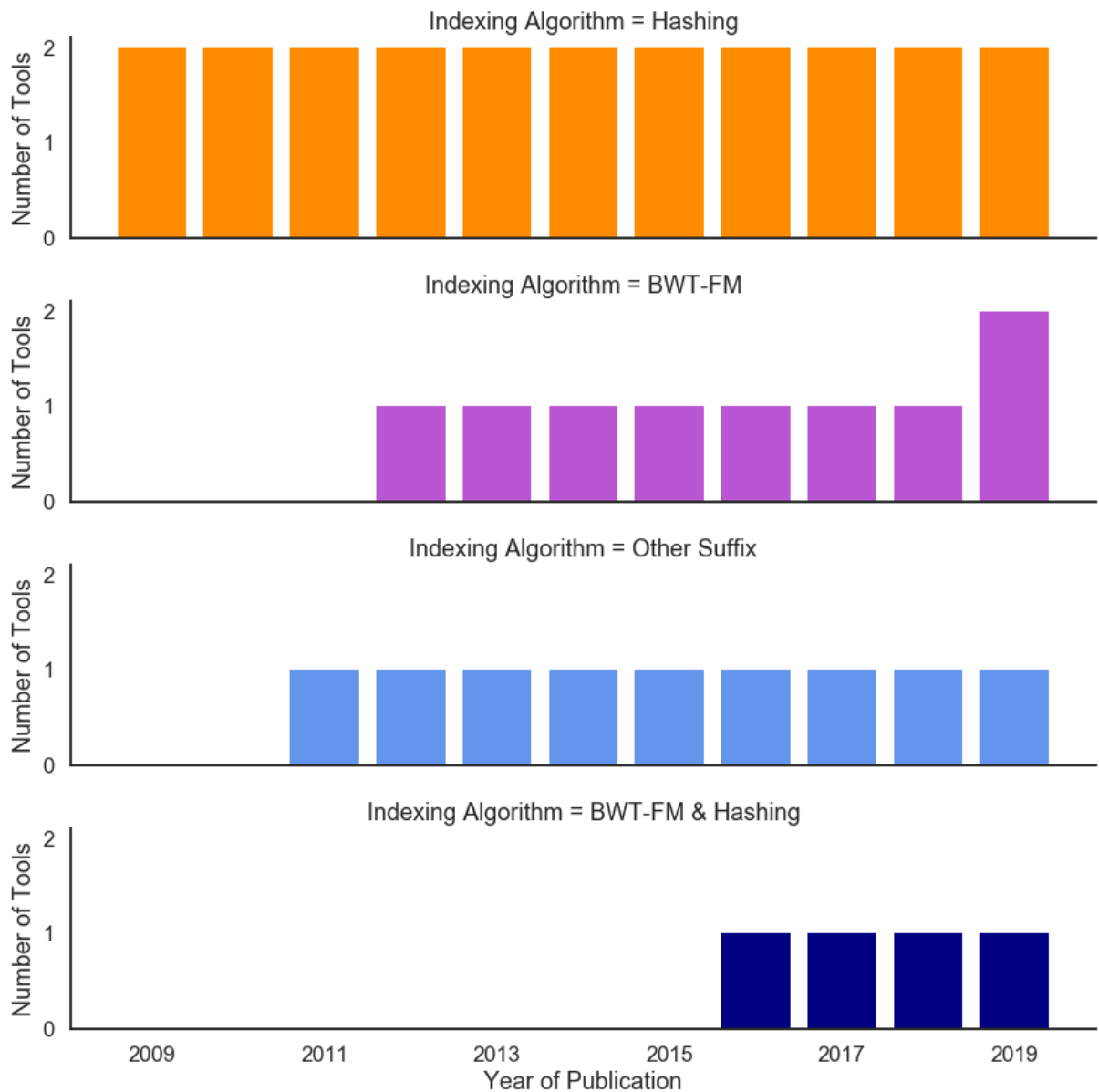

**Supplementary Figure 11. Histogram showing the cumulation of surveyed BS-Seq tools over time separated by the algorithm used for genome indexing.** Only stand-alone BS-Seq aligners tools are included (not the wrappers of existing DNA-Seq aligners).

**Supplementary Note 1**

We evaluate the effect of indexing on the end-to-end execution time of today's read alignment algorithms. We align a single read of length 100 bp to the human reference genome (hg38) using BWA-MEM (with -*a* parameter selected to report all mapping locations). Building the index for the human reference genome takes 3,476 seconds. The read alignment step using BWA-MEM takes only 3.4 seconds after building the index for the human reference genome. Now we want to perform brute-force read alignment for the same read sequence and the same reference genome that we use for the BWA-MEM experiment. We divide the human reference genome into about 3.3 billion sequences, each of which is 100 bp long. That is, the first sequence is the first 100 bp of the reference genome and the second sequence is the segment that starts from the second bp of the reference genome and ends at the 101th bp and so forth for the other sequences. We then use Edlib's global alignment tool (DP-based pairwise alignment) to check the similarity of the read sequence with each of the 3.3 billion generated sequences. We observe that Edlib takes about 24,200 seconds to complete the brute-force read alignment approach. This means that the indexing technique (and probably other filtering heuristics) used in BWA-MEM saves the execution time of read alignment by at least 7,100X. If we include the time needed to build the index in the total time of read alignment, then BWA-MEM is only 7X faster than the brute-force read alignment approach. Note that indexing the reference genome is performed only once for each reference genome.

**Supplementary Note 2**

We first built the index data, then ran the alignment procedure and extracted the data in bam format. Some tools do not provide the output in bam format, so in this case we used samtools toolkit to convert sam output to bam output.

To install samtools from conda: conda install -c bioconda samtools

1) Bowtie2

Build index:

bowtie2-build <reference_in> <index_basename>

*reference_fasta: fasta file of reference genome

*index_basename: write index data to files with this basename

Mapping WGS data:

bowtie2 -x <index_basename> -1 <r1_fastq> -2 <r2_fastq> | samtools view -bS - > output.bam

*r1_fastq, r2_fastq: fastq files of the paired end reads

2) Bowtie

Build index:

bowtie-build <reference_in> <index_basename>

*reference_in: fasta file of reference genome

*index_basename: write index data to files with this basename

Mapping WGS data:

bowtie -S <index_basename> -1 <r1_fastq> -2 <r2_fastq> | samtools view -bS - > output.bam

3) BWA

Build index:

bwa index <reference_fasta>

Mapping WGS data:

bwa mem <reference_fasta> <r1_fastq> <r2_fastq> | samtools view -bS - > output.bam

4) GSNAP

Build index:

gmap_build -D <destination_directory_path> -d <genome_name> <reference_fasta>

Mapping WGS data:

gsnap -D <destination_directory_path> -d <genome_name> <r1_fastq> <r2_fastq> -A sam |

samtools view -bS - > output.bam

5) HISAT2

Build index:

hisat2-build <reference_fasta> <index_basename>

Mapping WGS data:

hisat2 -q -x <index_basename> -1 <r1_fastq> -2 <r2_fastq> | samtools view -bS - > output.bam

*-q: input as fastq file

6) LAST

Build index:

lastdb -uNEAR -R01 <index_basename> <reference_fasta>

*-uNEAR and -R01 optional

Mapping WGS data:

lastal -Q1 <index_basename> <r1_fastq> <r2_fastq> | last-split > output.maf

*Q1: fastq-sanger format

7) minimap2

Build index:

Minimap2 -d <index_file> <reference_fasta>

* index file with ".mmi" extension

Mapping WGS data:

Minimap2 -a <index_file> <r1_fastq> | samtools view -bS - > output.bam

8) RMAP

rmap <read_fastq> -c <reference_fasta> -o output.sam | samtools view -bS - > output.bam

9) SMALT

Build index:

smalt index [options] <index_name> <reference_fasta>

Mapping WGS data:

smalt map <index_name> <r1_fastq> <r2_fastq> | samtools view -bS - > output.bam

10) SNAP

Build index:

snap-aligner <index_name> <reference_fasta> <index_dir_name>

Mapping WGS data:

snap-aligner paired <index_dir_name> <r1_fastq> <r2_fastq> -o output.bam

11) Subread

Build index:

subread-buildindex -o <index_name> <reference_fasta>

Mappins WGS data:

subread-align -t 1 -i <index_name> -r <r1_fastq> -R <r1_fastq> -o output.bam

**Supplementary Note 3**

To obtain the nucleotide count in all bacterial genomes possessed by NCBI, we utilized the tool RepoPhlAn(https://bitbucket.org/nsegata/repophlan) to download via ftp.ncbi.nih.gov all genomes contained in the genomes/all subdirectory. Taxonomic identifiers were used to identify bacterial genomes and subsequently obtain a nucleotide count.

```
# obtain RepoPhlAn
wget https://bitbucket.org/nsegata/repophlan/get/03f614c13cf0.zip
unzip 03f614c13cf0.zip
cd 03f614c13cf0
# run RepoPhlAn
./run.sh # this can take upwards of 5 days to complete this step
cd out/microbes_<time_stamp>/fna
# count number of bacterial nucleotides
nohup ls -U | xargs -P 15 -I{} sh -c "bzcat {} | grep -v '>'| wc -m " | awk '{sum+=$1}END{print sum}' > ~/bacteria_bp_count.txt
# 676153484835
```

The human genome build GRCh38 was obtained from NCBI via ftp and nucleotides counted in the following way:

```
# download the human genome
wget -r https://ftp.ncbi.nih.gov/genomes/Homo_sapiens/Assembled_chromosomes/seq/*
# select just the fasta files
```

```
cd ftp.ncbi.nih.gov/genomes/Homo_sapiens/Assembled_chromosomes/seq/

ls | grep -v"\.fa\." | xargs -I{} rm {}

#Uncompress

ls | xargs -I{} gunzip {}

# count nucleotides

ls *.fa | xargs -I{} sh -c " grep -v '>' {} | wc -m" | awk '{sum+=$1}END{print sum}' >
~/human_bp_count.txt

#3303852965

# compare the two

echo "`cat ~/bacteria_bp_count.txt` / `cat ~/human_bp_count.txt`" | bc -l

204.65604613702898246260
```

**Supplementary Materials**

**Install the read alignment tools**

We have selected tools available on bioconda and have installed them using the following commands (Table S1) .

**Public Sequence Data**

We used 9 WGS dataset for comparing the tools listed in Table S1.

The SRA ids of the 9 dataset are the following: ERR009309, ERR013138, ERR045708, ERR050158, ERR162843, ERR183377, SRR061640, SRR360549.

To download this data we used the sra toolkit which is available from conda package.

Here are the commands that we used for the downloading process:

To download sra toolkit: conda install -c bioconda sra-tools

To download fastq files:

 For single end fastq files: fastq-dump <SRA_id>

For paired end fastq files: fastq-dump --split-files <SRA_id>

**Compare the performance of the read alignments**

We recorded CPU time and RAM usage to compare the read alignment tools. Tools are ran in the UCLA's Shared Hoffman2 Cluster.

Here is the command that we used to submit our jobs in the cluster:

qsub -o <logfiles/> -e <logfiles/> -m bea -cwd -V -N <name_job> -l

h_data=32G,highp,time=24:00:00 <exe_script>

*-m bea: define mailing rules

- b- start time of the job

- e- end time of the job

- a- time when the job is aborted

-cwd: changes the directory to where your executed file is, all log output will be created in this

file unless you specify another directory (see command above output logs and error logs are

directed to a folder named logfiles)

-V: export environment variables

-N: give a name to the submitted job

-l h_date: resource allocation

-l highp: submission of high priority jobs

-l time: job running time

**Statistical analyses**

We model expected CPU time $c_{ij}$ across all algorithms $i$ and datasets $j$ using the following

gamma generalized linear mixed model regression

$log(E(c_{ij})) = \alpha + a_j + \beta_1 \ x \ Chain\_of\_seeds_{ij} + \beta_2 \ x \ Indexing_{ij} + \beta_3 \ x \ Year\_of\_publication_{ij} + \beta'_4 \ x \ Pairwise\_alignment_{ij}$ (1)

where $\alpha$ is the intercept and $a_j \sim N(0,\sigma_j)$ is a data-level random intercept modelling the shared noise within each data set. $\beta_1$ is the effect of the *Chain_of_seeds* where *Chain_of_seeds* is coded as zero for no and one for yes. $\beta_2$ is the effect of Indexing, where *Indexing* is coded as 0 for BWT-FM and 1 for hashing or suffix array, depending on the group being compared to BWT-FM. $\beta_3$ is the effect of *Year_of_publication*, coded as a continuous variable of year scaled to have mean zero and variance one. $\beta_4$ is a vector with the effects of *Pairwise_alignment*, where *Pairwise_alignment* is a matrix of indicator variables for HD, Non-DP Heuristic, and SW algorithms, making NW the reference category. Parameter estimates are provided in (Table S3). We use a likelihood ratio test to test the effect of each variable discussed, e.g. year of publication or indexing, on the CPU time.

We use a similar model for the median across all datasets of the expected RAM usage med_$mem_i$, i.e.

$log(E(\mathrm{med}\_mem_i)) = \alpha + \beta_1 \ x \ Chain\_of\_seeds_{ij} + \beta_2 \ x \ Indexing_{ij} + \beta_3 \ x \ Year\_of\_publication_{ij} + \beta_4 \ x \ Pairwise\_alignment_{ij}$ (2)

Parameter estimates are provided in (Table S4). Note that, as memory usage does not vary considerably within algorithms across data sets, we use the median expected RAM usage across all datasets for each algorithm.

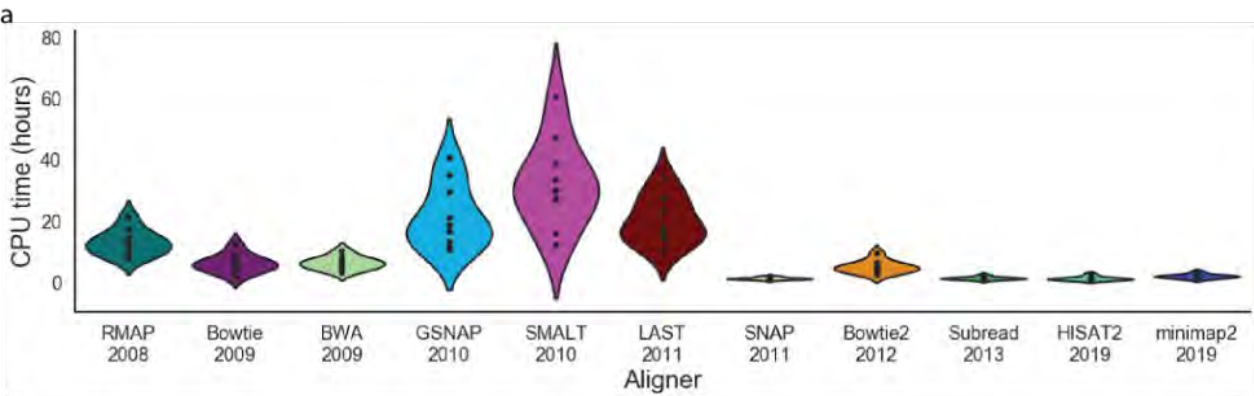

a

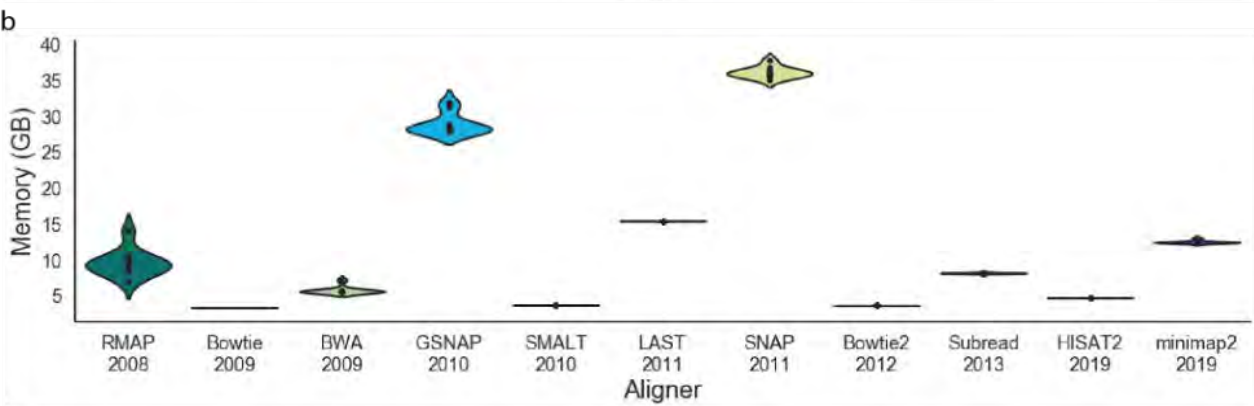

b

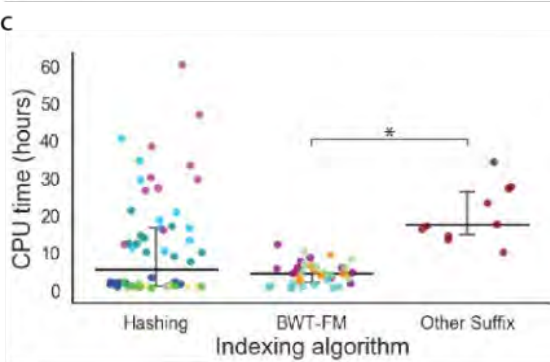

c

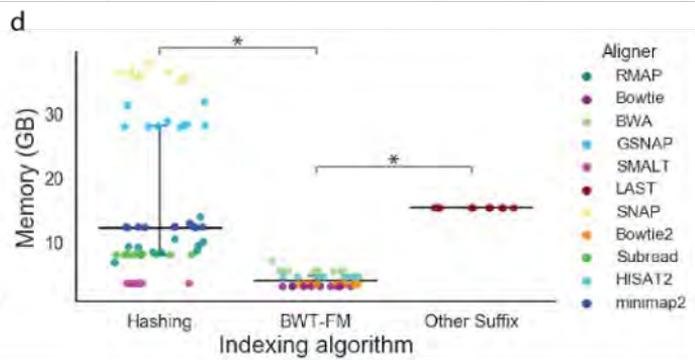

d

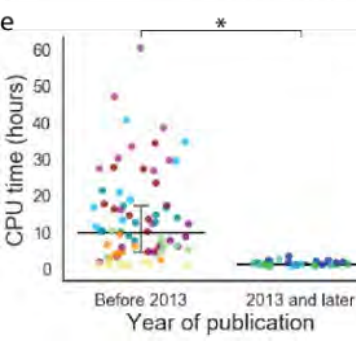

e

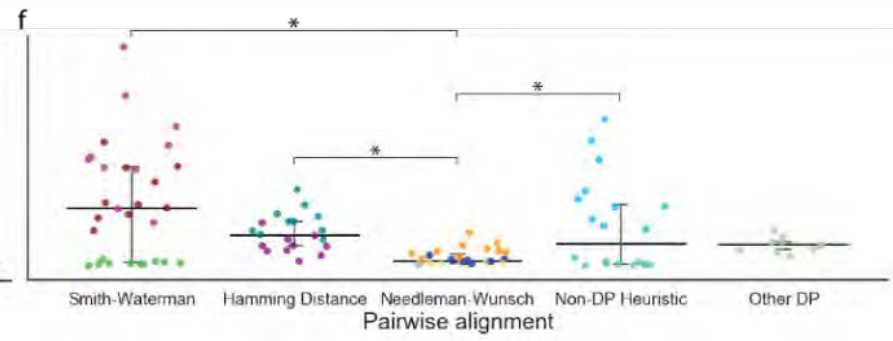

f